\documentclass[twocolumn]{aastex631}

\newcommand{\mj}{M$_{Jup}$}
\newcommand{\mum}{$\mu$m}
\newcommand{\hd}{HD~80606}
\newcommand{\hdb}{HD~80606~b}
\newcommand{\petit}{\texttt{petitRADTRANS}}

\newcommand{\teff}{T$_{\rm eff}$}

\newcommand{\kms}{$\rm km\,s^{-1}$}
\newcommand{\kpvsys}{$K_{\rm p} - \Delta v_{\rm sys}$}
\newcommand{\kp}{$K_{\rm p}$}
\newcommand{\dvsys}{$\Delta v_{\rm sys}$}
\newcommand{\ucla}{Department of Physics \& Astronomy, 430 Portola Plaza, University of California, Los Angeles, CA 90095, USA}
\newcommand{\jpl}{Jet Propulsion Laboratory, California Institute of Technology, 4800 Oak Grove Dr.,Pasadena, CA 91109, USA}
\newcommand{\ipac}{IPAC, Mail Code 100-22, Caltech, 1200 E. California Blvd., Pasadena, CA 91125, USA}

\shorttitle{HD 80606 b with NIRSPEC}
\shortauthors{Finnerty et al.}
\graphicspath{{./}{figures/}}

\begin{document}

\title{Limits On the Post-eclipse Emission Spectrum of HD~80606~b From High-Resolution Spectroscopy}

\author[0000-0002-0786-7307]{Luke Finnerty}
\affiliation{\ucla}

\author{Aurora Kesseli}
\affiliation{\ipac}

\author{Kyle Pearson}
\affiliation{\jpl}

\author[0000-0002-5627-5471]{Charles Beichman}
\affiliation{\jpl}

\author{Michael P. Fitzgerald}
\affiliation{\ucla}

\begin{abstract}

We present Keck/NIRSPEC $K$-band observations of \hdb, one of the most eccentric known exoplanets. \hdb\ was observed after secondary eclipse, close to periastron, when the planet passes within 0.03 AU of \hd\ and the rapid heating of the atmosphere may lead to extreme chemical changes and a temporary thermal inversion. The rapid change in the planetary radial velocity near periastron is sufficient to enable high-resolution cross-correlation spectroscopy (HRCCS) analysis, which produces a tentative detection ($\rm SNR\sim4$) of \hdb. Injection-recovery tests appear to reject strong thermal inversions near periastron, consistent with recent results from \textit{JWST}. We also perform atmospheric retrievals with free parameters for the Pressure-Temperature ($P-T$) profile and with a profile matched to the \textit{JWST} results, which suggest the presence of absorption features from CH$_4$ and CO. While \hdb\ is not definitively detected in these data, these results are consistent with \textit{JWST} observations, which found the post-eclipse atmosphere of \hdb\ shows weak absorption features from these species. Future observations with higher spectral resolution and/or wider wavelength coverage are needed for a confident atmospheric detection of \hdb\ via high-resolution spectroscopy alone, but such observations are a challenge to schedule due to the 111-day orbital period. 

\end{abstract}

\section{Introduction}\label{sec:intro}

For many years \hdb\ held the record for the  most highly eccentric planet. Discovered by the radial velocity (RV) technique in 2001 \citep{naef2001}, \hdb\ has a mass of 4.1~\mj, an orbital period of 111.4~days and an eccentricity of $\epsilon$=0.93. System properties are summarized in Table \ref{tab:props}. \hdb\ continues to be compelling for further study as it was discovered by Spitzer to be eclipsing \citep{laughlin2009}, passing within 0.03~AU  of its host G5V star. During its rapid periastron passage of a few tens of hours, the insolation increases a thousand-fold and the equilibrium temperature rises from 400 K to 1400 K (Figure~\ref{fig:temps}a)  providing a unique opportunity to explore the dynamical response of an atmosphere under an extreme external forcing function. Spitzer's photometric observations of eclipses in 2009 and 2010 at 8.0 and 4.5~\mum, respectively, were used to infer timescales for radiative, dynamical, and chemical processes \citep{dewit2016, lewis2017}.  

Two approved JWST Cycle 1 programs have or will make spectroscopic observations around the eclipse of \hdb\  (which is conveniently oriented a few hours before periapsis) and are poised to exploit this highly variable environment to study a wide variety of atmospheric properties, including composition, chemical and dynamical timescales, and large scale atmospheric motions. 

The JWST observations to data have covered two separate eclipses. MIRI LRS (5-12 $\mu$m \& R$\sim$ 40-150) observations obtained on 2025-April-11 (T. Kataria, PI; PID\#2008) will constrain the internal heating of HD 80606b, probe variations in clouds and chemistry throughout \hdb’s orbit, and assess the role of dynamical mixing in \hdb’s atmosphere. A second team  (J. Sikora, PI; PID\#2488) used  NIRSpec (3-5 $\mu$m \& R$\sim$2700) on 2022-Nov-1 to explore the formation and evolution of atmospheric clouds, which are expected to vary rapidly due to evaporation \& sublimation. 

This paper reports results from  a well placed eclipse on  2024-Jan-21 (UT) using NIRSPEC in K-band on the Keck-II telescope to complement the JWST observations with new wavelengths obtained with much higher spectral resolution (R$\sim$30,000). 


\begin{deluxetable}{ccc}
\tabletypesize{\scriptsize}
    \tablehead{\colhead{Property} & \colhead{Value} & \colhead{Ref.}}
    \startdata
        & \textbf{\hd} & \\
        \hline
        RA & 09:22:37.67 &  \citet{gaiaedr3} \\
        Dec & +50:36:13.6 &  \citet{gaiaedr3} \\
        Sp. Type & G5 &  \\
        $K_{\rm mag}$ & $7.32\pm0.02$ & \citet{cutri2003} \\
        $K_s$ [$\rm m\,s^{-1}$] & $469.22\pm0.61$ & \citet{pearson2022} \\
        Mass [$\rm\ M_\odot$] & 1.05 & \citet{pearson2022}   \\
        Radius [$\rm\ R_\odot$] & $1.05\pm0.01$  & \citet{pearson2022} \\
        \teff\ [K] & $5565\pm92$ & \citet{rosenthal2021} \\
        $\log g$ [cgs] & $4.4\pm0.03$ & \citet{rosenthal2021}   \\
        $v\sin i$ [\kms] & $1.8\pm0.5$ & \citet{bonomo2017} \\
        $v_{\rm rad}$ [\kms] & $4.16\pm0.37$ & \citet{gaiaedr3} \\
        $\rm [Fe/H]$ & $0.36\pm0.18$ &  \citet{hypatia} \\
        $\rm [C/H]$ & $0.33\pm0.10$ & \citet{hypatia} \\
        $\rm [O/H]$ & $0.19\pm0.11$ &  \citet{hypatia} \\
        \smallskip \\
        \hline
         & \textbf{\hdb} & \\
        \hline
        Period [day] &  $111.436765\pm0.000074$  & \citet{pearson2022} \\
        $\rm T_{\rm transit}$ [BJD] & $2458888.07466\pm0.00204$  & \citet{pearson2022}  \\
        $\rm T_{\rm peri}$  [BJD] & $2458882.344\pm0.0021$ & \citet{pearson2022}\\
        $a$ [AU] & $0.4603\pm0.0021$ & \citet{pearson2022}   \\
        $e$ & $0.93183\pm0.00014$ & \citet{pearson2022}  \\
        $\omega$ & $-58.887^\circ\pm0.043$ & \citet{pearson2022} \\
        $i$ & $89.24^\circ\pm0.01$ & \citet{pearson2022} \\
        $K_{\rm p}$ [\kms] & 123.8  & Calc.  \\
        Mass [$\rm M_J$] & $4.1641\pm0.0047$ & \citet{pearson2022}  \\
        Radius [$\rm R_J$] & $1.032\pm0.015$ & \citet{pearson2022} \\
    \enddata
    \caption{Stellar and planetary properties for the \hd\ system.} 
    \label{tab:props}
\end{deluxetable}

\begin{figure*}[t!]
 \centering
 \includegraphics[width=0.85\textwidth]{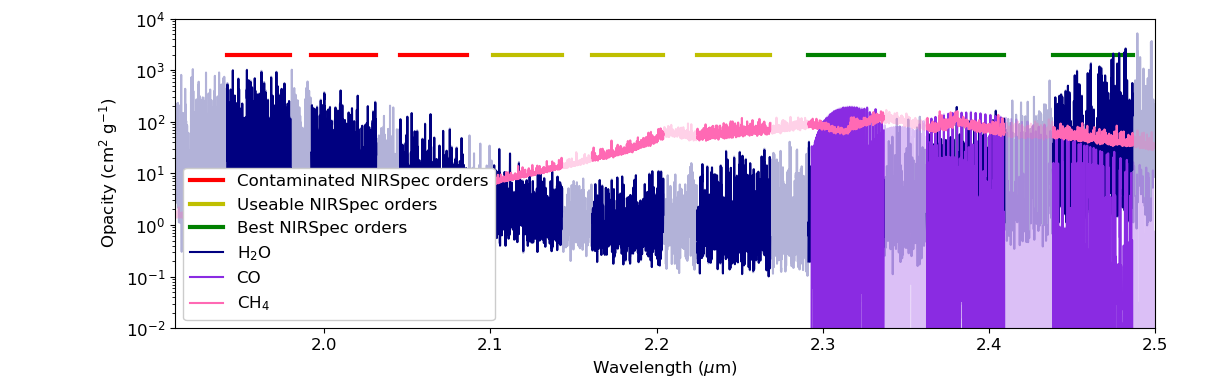}
 \caption{Opacities of the dominant species (H$_2$O, CO, and CH$_4$) in the NIRSPEC K-band spectral region. NIRSPEC contains 9 spectral orders in the K-band region, 6 of which can be calibrated and used in our analysis (3 are highly contaminated). H$_2$O, CO, and CH$_4$ all contain major bandheads within this spectral region and so we expect to be able to detect them, if they are present. \label{fig:opacs}}
 \end{figure*}

In addition to exploring a different wavelength range from JWST (Figure~\ref{fig:opacs}), NIRSPEC's higher spectral resolution has two major advantages compared with JWST's lower resolution, higher sensitivity data. First, high resolution spectroscopy has been very successful at unambiguously identifying (or excluding) key carbon- and oxygen-bearing species in exoplanet atmospheres by resolving large numbers of individual molecular absorption or emission lines \citep[e.g.][]{birkby2018, snellen2025}. In contrast, overlapping spectral features from different molecules or narrow-line absorption from molecules such as CO can make unambiguous identification of some molecules difficult with low-resolution spectroscopy \citep{welbanks2025}. Keck II/NIRSPEC, as the spectrometer used for KPIC, has a proven track record of detecting and characterizing exoplanet atmospheres via high-resolution spectroscopy \citep{finnerty2023, finnerty2024, finnerty2025a, finnerty2025b, finnerty2025c}

Secondly, high spectral resolution observations can also resolve atmospheric dynamics such as winds and planet rotation \citep[e.g.][]{snellen2010, ehrenreich2020, kesseli2022, lesjak2023}. In the case of \hdb, the precise ephemerides from \citet{pearson2022} can be used to precisely calculate the planet's velocity and provide a basis for measuring excess red or blueshifts of the planet signal in emission. Tidally locked hot Jupiters on average show an excess net blueshift of $2-5$ \kms, which is caused by atmospheric dynamics and attributed to a global day-to-night wind in the atmospheres \citep[e.g.][]{snellen2010}. Measuring velocity shifts in \hdb\ would help to constrain global circulation models (GCMs) for eccentric hot planets. 

\section{Observations and Data Reduction}
\subsection{Observations}

We observed \hd\ using Keck II/NIRSPEC \citep{nirspec, nirspecupgrade, nirspecupgrade2} on 2024 January 21 from 6:16 to 15:20 UT. We had originally proposed to perform these observations using Keck II/KPIC \citep{kpic, jovanovic2025}, but switched to the seeing-limited mode of NIRSPEC due to instrument scheduling conflicts. As NIRSPEC is the spectrometer used in KPIC, this does not impact the wavelength coverage or spectral resolution of the observations. Using the slit-fed mode results in a significant ($\sim6\times$) throughput increase compared with KPIC, at the cost of reduced wavelength/blaze stability and increased sky background.

The observations included a complete secondary eclipse from approximately 8:30 to 10:30 UT. The pre-eclipse observations were taken at high airmass (2.83--1.52), leading to significantly worse signal-to-noise (SNR) compared with the post-eclipse observations, which ranged in airmass from 1.16--1.4 as \hd\ set. High clouds caused significant ($>2$ mag) extinction beginning around 14:20 UT. All exposures were taken with a 60 second integration time with \hd\ in the center of the $0.288'' \times 24''$ slit and the Kband-new filter.

\begin{figure*}
     \centering
     \includegraphics[width=0.45\linewidth]{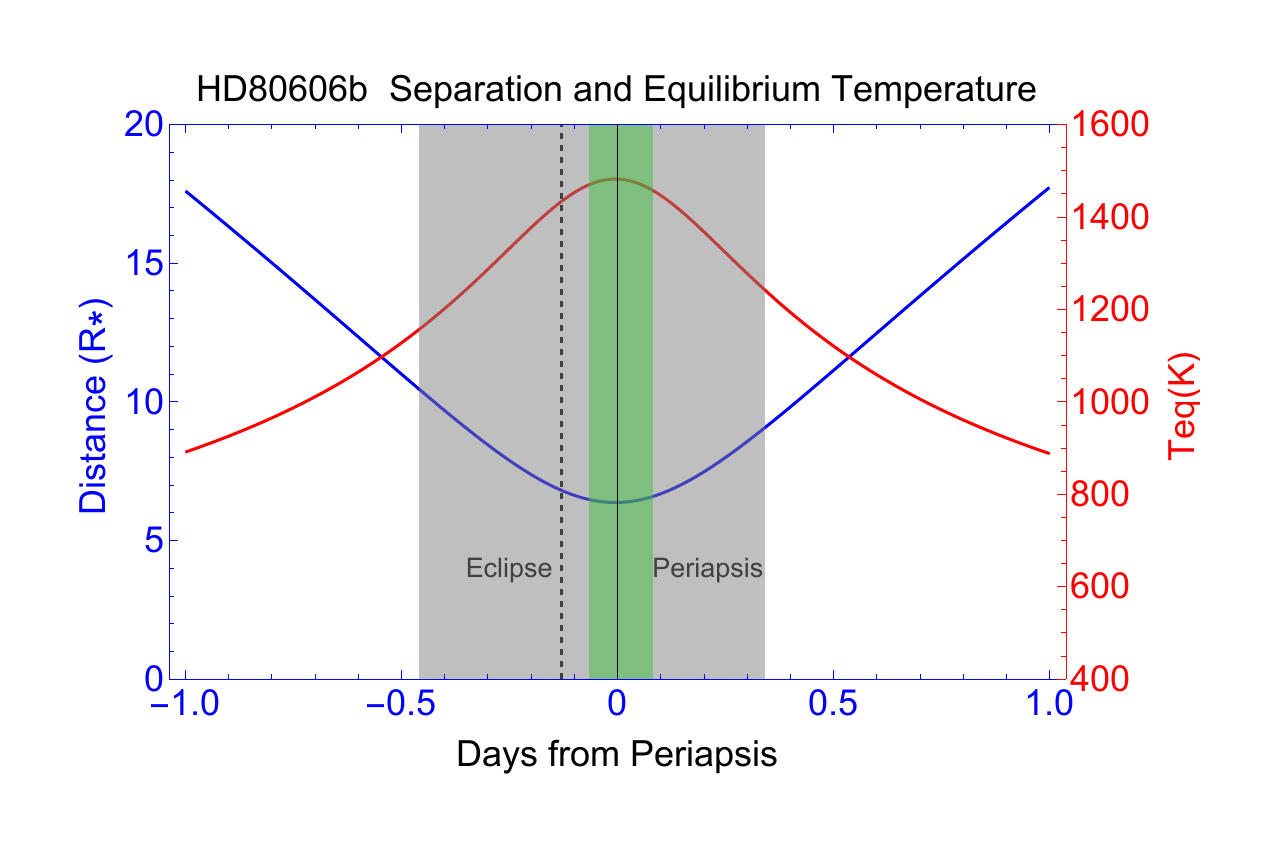}\includegraphics[width=0.45\linewidth]{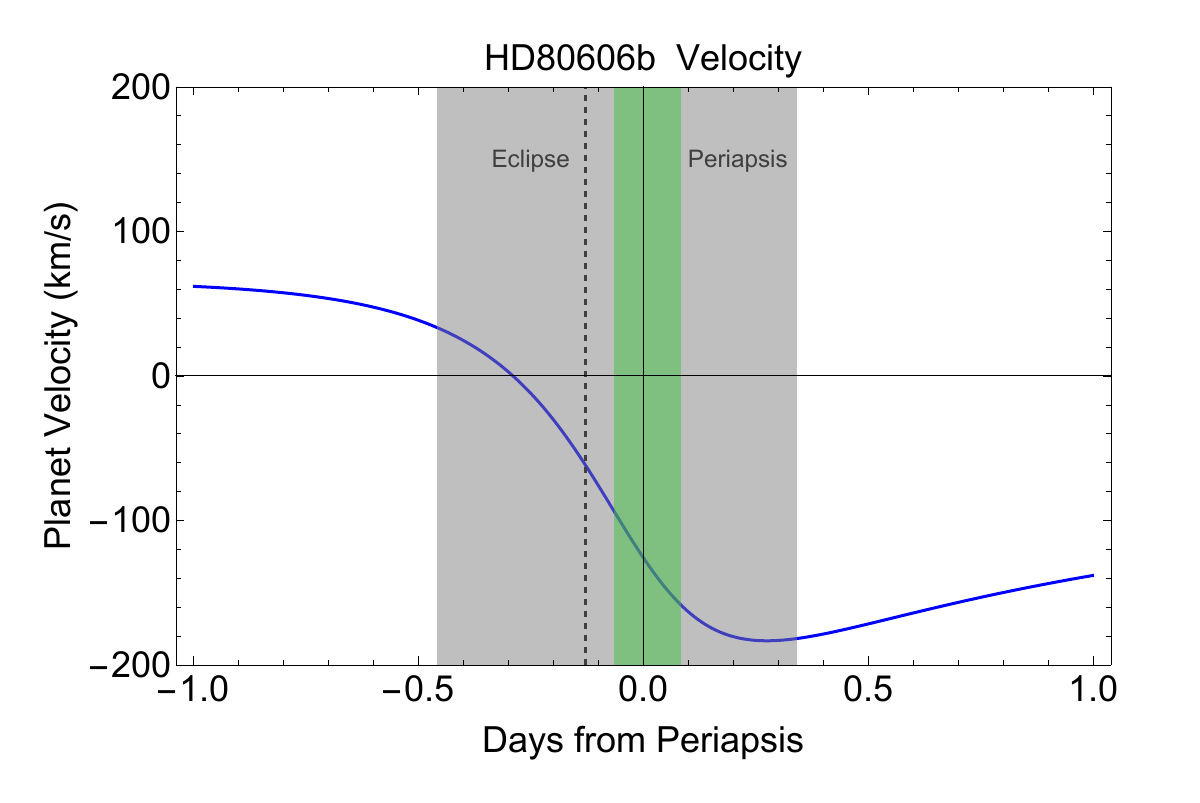}
     \caption{(left) The blue curve shows the separation of \hdb\ from its host star in units of the stellar radius as a function of time  immediately before and after periapsis. The eclipse midpoint is a few hours earlier and is denoted by a dashed black line. The red curve is an illustrative equilibrium temperature for the planet driven only by instantaneous insolation based on the square root of its distance from the star consistent with the observations of \citep{laughlin2009}. The gray  shaded region shows the coverage of the \citet{sikora2024} \textit{JWST} observations, and the green  shaded region shows the approximate times of the NIRSPEC observations presented here. right) The velocity of the planet in km s$^{-1}$. Values  are derived from the model and parameters described in \citet{pearson2022}.
     \label{fig:temps}}
 \end{figure*}

As a result of the high airmass negatively impacting the pre-eclipse frames, we restrict our calculation of the log-likelihood to the 180 post-eclipse frames obtained between 10:45 UT and 14:20 UT, when clouds began to significantly impact the data quality. We coadd pairs of consecutive frames to reduce the data volume for the analysis, and to improve the quality of the detrending, resulting in a 90 frame time series from the original 180 raw science frames.The phase, temperature, and velocity coverage of these observations is illustrated in Figure \ref{fig:temps}.

\subsection{Data Reduction}

The data were reduced using a modified version of the KPIC DRP\footnote{\href{https://github.com/kpicteam/kpic_pipeline/}{https://github.com/kpicteam/kpic\_pipeline/}} as described in \citet{finnerty2025a, finnerty2025b}, with minor alterations to address the increase in sky background when using slit-fed NIRSPEC compared with KPIC. While the KPIC fibers are fixed in the NIRSPEC slit, in slit-fed mode the science target can move along the slit over the observation sequence, leading to potential variations in the blaze function and wavelength solution from frame to frame. The slit also results in significantly greater sky background compared to the fiber feed, including noticeable telluric emission features. To minimize the wavelength and blaze variations, we obtained our observations in staring mode. Dark current was removed using afternoon dark frames. We do not perform additional out-of-trace subtraction of for telluric emission features, which are instead removed through the median division and PCA described in Section \ref{ssec:detrend}

We initially attempted to use the Gaussian-Hermite trace profile/line-spread function (LSF) fitting routines previously described in \citet{finnerty2025a, finnerty2025b, finnerty2025c} without modification. However, the resulting LSF FWHM was substantially broader than expected, giving a spectral resolution $R = \lambda/\Delta\lambda\sim30,000$ compared with the expected value $R\sim35,000$ from previous observations with KPIC and the Keck/NIRSPEC documentation. This fitting procedure is applied to a stack of science frames in order to maximize signal-to-noise. Stacking slit-fed frames will broaden the trace profile in the stack compared to individual images due to jitter along the direction of the slit, and this broadening results in a lower measured spectral resolution.

We avoided this broadening by first computing the shift from the nominal trace center for each frame, then fitting the trace profile with these per-frame offsets applied. This resulted in significantly better flux extraction, particularly towards the end of the observation sequence as the star began to move significantly more along the slit ($\sim 6$ pixels compared with $\sim1$ pixel at the start of the observation sequence). As the resulting profile widths were still broader than expected for the instrument, we modeled the LSF using a Gaussian with a standard deviation of 1.3 pixels, which was used for KPIC observations of hot Jupiters before the more sophisticated LSF fitting procedure was implemented \citep{finnerty2023}. We prefer to possibly under-estimate the LSF width rather than over-estimate because we include a rotational broadening kernel in the retrievals, which can broaden the planet lines to match the true LSF.

Similar to the KPIC observing procedure \citep[e.g.][]{finnerty2023}, we observed several late-type giant stars for wavelength calibration. We then fit these observations with a PHOENIX stellar model \citep{phoenix} and a telluric model from the Planetary Spectrum Generator \citep{psg} to derive a wavelength solution. Compared with a telluric-only approach to deriving the wavelength solution, using observations of an M-type giant greatly increases the number of lines used for fitting due to absorption features in the stellar spectrum, particularly CO. \citep{horstman2024} find this procedure yields a final radial velocity precision of $0.4-1.0$ \kms.

However, because the wavelength solution varies along the NIRSPEC slit, we used these targets only for a rough calibration. We then used the resulting wavelength solution as a starting point to refit a final wavelength solution to the coadded spectrum of \hd. This process resulted in a good wavelength solution for all 9 orders on the NIRSPEC detector. However, orders 37--39 (1.94-2.09 $\mu$m) are heavily contaminated by telluric absorption features and omitted from subsequent analysis, leaving orders 31-36 covering 2.1-2.48 $\mu$m, consistent with previous results from KPIC \citep[e.g.][]{finnerty2025c}. 

The frame-to-frame motion along the slit will also cause a shift to the wavelength solution between each frame. Comparing previous KPIC observations (which use the same spectrometer) taken with multiple science fibers suggests the NIRSPEC wavelength solution shifts by approximately 0.01\r{A}/pixel along the slit in the $K$ band, while the width of the resolution element is approximately 0.6 \r{A} at 2.3 \micron. The shift in the wavelength solution associated with a $\sim0.5$ pixel jitter along the slit axis is therefore negligible ($<1\%$ of a resolution element), and even a large $\sim5$ pixel offset would not substantially shift the wavelength solution. Inspection of the spectral time series does not reveal any significant frame-to-frame changes in the wavelength solution. We therefore use a single wavelength solution from fitting a stellar $\times$ telluric model to the full stack of science frames.

\subsection{Detrending}\label{ssec:detrend}

Despite concerns about frame-to-frame blaze function variations, we find that the detrending procedure described in \citet{finnerty2025c} is sufficient to remove systematics from the time series. This procedure consists first scaling all spectra in the time series to a consistent continuum level, then masking the first/last 50 wavelength channels of each order, followed by any channels with $<70\%$ telluric transmission, and finally masking the 3\% of remaining channels with the highest variance. This is then followed by a 6$\times$ median absolute deviation clip and then dropping 4-8 Principal Components (PCs) computed over the time of the spectral time series for each order. The omitted principal components are saved and added to the forward model in order to replicate the distortions to the underlying planet signal caused by PCA \citep{line2021}. This procedure is identical to that described in previous KPIC HRCCS analyses \citep{finnerty2025a, finnerty2025b, finnerty2025c}. We find that the number of omitted PCs has some impact on the cross-correlation and retrieval results, which we discuss below.

\subsection{Retrieval analysis}

\begin{deluxetable*}{ccccc }
    \tablehead{\colhead{Name} & Symbol  & Prior & \colhead{Max-L} & \colhead{Median}}
    \startdata
         & \textbf{Fixed $P-T$ profile} & \\
        \hline
        $K_{\rm p}$  [\kms] & $\Delta K_p$ & Uniform(-50,50) & $48.0$ & $-0.0^{+34.0}_{-30.0}$ \\
        $v_{\rm sys}$ offset [\kms] & $\Delta v_{sys}$ & Uniform(-30,30) &  $9.0$ & $-2.0^{+18.0}_{-18.0}$ \\
        Rotational velocity [\kms] & $v_{rot}$ & Uniform(0,15) &  $2.3$ & $6.7^{+5.0}_{-4.3}$ \\
        log H$_2$O mass-mixing ratio & log H$_2$O  & Uniform(-12,-0.3) & $-3.9$ & $-5.8^{+3.2}_{-4.1}$ \\
        log CO mass-mixing ratio & log CO & Uniform(-12,-0.3) &  $-1.8$ & $-3.8^{+1.7}_{-5.1}$ \\
        log CH$_4$ mass-mixing ratio & log CH$_4$  & Uniform(-12,-1) &   $-4.8$ & $-7.5^{+3.3}_{-2.9}$ \\
        log H$_2$ mass-mixing ratio & $\log \rm H_2$ & Uniform(-0.4,-0.05) & $-0.4$ & $-0.2^{+0.1}_{-0.1}$ \\
        log scale factor & $\log a$ & LogNormal(0,0.1) & $0.19$ & $0.02^{+0.09}_{-0.1}$ \\ \\
        \hline
        & \textbf{Free $P-T$ profile} & \\
        \hline
        log infrared opacity [$\rm cm^{2} g^{-1}$] & $\log \kappa$ & Uniform(-4,2) & $1.2$ & $-0.3^{+1.2}_{-2.0}$ \\
        log infrared/optical opacity & $\log \gamma$  & Uniform(-3,3) &  $-1.2$ & $-1.3^{+1.7}_{-0.8}$ \\
        Intrinsic Temperature [K] & $\rm T_{int}$ & Uniform(10,300) &   $123$ & $126.0^{+98}_{-75}$ \\
        Equilibrium temperature [K] & $\rm T_{equ}$ & Uniform(400,2000) &   $1940$ & $1160^{+510}_{-430}$ \\
        $K_{\rm p}$  [\kms] & $ K_{\rm p}$  & Uniform(-50,50) & $-23.0$ & $-10.0^{+42.0}_{-16.0}$ \\
        $v_{\rm sys}$ offset [\kms] & $\Delta v_{\rm sys}$ & Uniform(-30,30) & $-11.0$ & $-3.0^{+16.0}_{-12.0}$ \\
        Rotational velocity [\kms] & $v_{\rm rot}$ & Uniform(0,15) &  $1.5$ & $7.5^{+5.1}_{-4.6}$ \\
        log H$_2$O mass-mixing ratio & log H$_2$O  & Uniform(-12,-0.3) & $-6.6$ & $-7.4^{+4.2}_{-2.9}$ \\
        log CO mass-mixing ratio & log CO & Uniform(-12,-0.3) & $-1.4$ & $-3.0^{+1.7}_{-5.0}$ \\
        log CH$_4$ mass-mixing ratio & log CH$_4$  & Uniform(-12,-1) & $-1.2$ & $-3.5^{+1.4}_{-4.2}$ \\
        log H$_2$ mass-mixing ratio & $\log \rm H_2$ & Uniform(-0.4,-0.05) & $-0.2$ & $-0.2^{+0.1}_{-0.1}$ \\
        log scale factor & $\log a$ & LogNormal(0,0.1) & $0.15$ & $0.01^{+0.09}_{-0.09}$ \\
        \hline
        & \textbf{Free $P-T$ profile, no inversions} & \\
        \hline
        log infrared opacity [$\rm cm^{2} g^{-1}$] & $\log \kappa$ & Uniform(-4,2) &$0.4$ & $-0.2^{+1.1}_{-1.7}$ \\
        log infrared/optical opacity & $\log \gamma$  & Uniform(-3,0) &  $-1.3$ & $-1.4^{+0.6}_{-0.7}$ \\
        Intrinsic Temperature [K] & $\rm T_{int}$ & Uniform(10,300) &   $98$ & $127^{+97}_{-74}$ \\
        Equilibrium temperature [K] & $\rm T_{equ}$ & Uniform(400,2000) &  $1910$ & $1220^{+480}_{-430}$ \\
        $K_{\rm p}$  [\kms] & $ K_{\rm p}$  & Uniform(-50,50) & $-24.0$ & $-16.0^{+42.0}_{-13.0}$ \\
        $v_{\rm sys}$ offset [\kms] & $\Delta v_{\rm sys}$ & Uniform(-30,30) & $-11.0$ & $-5.0^{+14.0}_{-11.0}$ \\
        Rotational velocity [\kms] & $v_{\rm rot}$ & Uniform(0,15) &  $1.4$ & $7.6^{+4.9}_{-4.5}$ \\
        log H$_2$O mass-mixing ratio & log H$_2$O  & Uniform(-12,-0.3) & $-6.4$ & $-7.4^{+4.2}_{-2.9}$ \\
        log CO mass-mixing ratio & log CO & Uniform(-12,-0.3) &  $-2.7$ & $-2.7^{+1.5}_{-4.3}$ \\
        log CH$_4$ mass-mixing ratio & log CH$_4$  & Uniform(-12,-1) &  $-2.4$ & $-3.4^{+1.3}_{-3.9}$ \\
        log H$_2$ mass-mixing ratio & $\log \rm H_2$ & Uniform(-0.4,-0.05) &  $-0.2$ & $-0.2^{+0.1}_{-0.1}$ \\
        log scale factor & $\log a$ & LogNormal(0,0.1) & $0.07$ & $0.02^{+0.09}_{-0.09}$ \\
    \enddata
    \caption{Retrieved parameters and priors for the three $P-T$ cases omitting 6 principal components. Additionally, we require the atmospheric temperature stay below 3500 K at all pressure levels. For nearly all parameters, the marginalized posterior spans the prior range. The full corner plots are shown in Appendix~\ref{app:corner}. }
    \label{tab:priors}
\end{deluxetable*}

We use the cross-correlation/retrieval analysis pipeline developed for the KPIC hot Jupiter survey \citep{finnerty2023, finnerty2024, finnerty2025a, finnerty2025b, finnerty2025c}, which uses \petit\ for radiative transfer calculations \citep{prt:2019, prt:2020, Nasedkin2024}.  We use a PHOENIX \citep{phoenix} model for the stellar spectrum with $\rm T = 5500$ K, $\log g = 4.5$, and $[\rm Fe/H] = +0.5$, which is the model in the PHOENIX library closest to the stellar parameters listed in Table \ref{tab:props}. This model is rotationally broadened to $v\sin i = 1.8$ \kms\ using the \citet{carvalho2023} method. For the model fitting, we use the \texttt{PyMultiNest} \citep{buchner2014} wrapper for \texttt{MultiNest} \citep{feroz2008, feroz2009, feroz2019} to perform Nested Sampling \citep{skilling2004}. Sampling was performed with 1200 live points to a convergence criteria $\Delta \log z < 0.01$. Retrievals were run for 4, 6, and 8 omitted principal components (PCs). For a given number of components, we ran two retrievals with different parameterizations/priors on the $P-T$ profile. The full set of retrieved parameters and priors is listed in Table \ref{tab:priors}.

For the Pressure-Temperature ($P-T$) profile, we perform two sets of retrievals using the \citet{guillot2010} parameterization, and another set using a fixed $P-T$ profile with parameters chosen to match the results of the \textit{JWST} analysis presented in \citet{sikora2024}. The $P-T$ parameters are the log infrared opacity ($\log \kappa$), log of the infrared/optical opacity ratio ($\log \gamma$), the intrinsic temperature ($\rm T_{int}$), and the equilibrium temperature ($\rm T_{equ}$). For the fixed $P-T$ profile, we adopted values $\log \kappa = -2.5$, $\log \gamma = -2.0$, $\rm T_{int} = 300\ K$, and $\rm T_{equ} = 1600\ K$, chosen to approximately match the profile \citet{sikora2024} reported at eclipse (see their Figure 8). For the retrievals including the $P-T$ profile parameters, we ran two sets, on with $\log \gamma < 0$ as a prior, which prevents thermal inversions, and one set with a more permissive prior allowing inverted or non-inverted $P-T$ profiles

Following \citet{finnerty2025c}, we do not include $\log g$ as a free parameter in the retrieval due to degeneracies with $\log \kappa$, which is a fundamental behavior of the $P-T$ implementation. Equation 29 of \citet{guillot2010}, which we use for the $P-T$ profile, parametrizes the atmospheric temperature in terms of the optical depth $\tau = \kappa_{th}m$, where $\kappa_{\rm th}$ is the thermal opacity and $m$ is the column mass. To compute the temperature at each pressure layer, \texttt{petitRADTRANS} uses ${\rm d}m = \rho {\rm d}z$, where ${\rm d}z$ is an incremental change in altitude. Assuming hydrostatic equilibrium gives ${\rm dP / d}z = -g\rho$, and therefore $m = {\rm P}/g$, and $\tau = \kappa_{\rm th}{\rm P}/g$. This $\tau$ is then used to compute $\rm T(P)$ via Equation 29 of \citet{guillot2010}. In this implementation, $\log g$ impacts the $P-T$ profile directly inverse to $\log \kappa$, while otherwise having only a weak effect on the shapes of planetary lines. This makes inclusion of both $\log g$ and $\log \kappa$ redundant given the quality of presently available data. 

Similarly, we do not include a cloud deck in our retrievals, as initial attempts including cloud parameters returned flat posteriors for the corresponding parameters. The loss of continuum information during the detrending process appears to make these data largely insensitive to the constant-opacity gray cloud model, and the lack of constraints on the cloud parameters therefore does not necessarily establish that the atmosphere of \hdb\ is cloud-free. Some combination of chromatic cloud models, higher-sensitivity and/or broader wavelength observations, and lower resolution observations which preserve the continuum level are needed to establish the presence and properties of clouds in \hdb. 

We include the planet radial velocity semiamplitude, \kp, and a systemic velocity offset, \dvsys, as free parameters. Given the precise ephemeris available for \hdb \citep{pearson2022}, we expect \kp\ to be close to the nominal value of 123.8 \kms\ and \dvsys$\sim0$. Small velocity offsets ($\lesssim10$ \kms) are consistent with wind speeds reported in UHJs \citep{wardenier2021, pai2022} and global circulation models \citep{wardenier2023, wardenier2025}, while larger offsets are suggestive of a non-planetary origin of those features. We also include a rotational broadening kernel as a free parameter. While we are skeptical of the physical significance of this parameter given the simple LSF treatment, we include it as a nuisance parameter.

We fit for fixed-with-altitude volume-mixing ratios of H$_2$O, CO, CH$_4$, and H$_2$, and fill the remainder of the atmosphere with He. For H$_2$O, we used the opacities based on the \citet{polyansky2018} POKAZATEL linelist. For CO, we used the HITEMP linelist \citep{hitemp2020}. For CH$_4$, we use the \citet{hargreaves2020} linelist. H$_2$ opacity is from \citet{rothman2013}. We also include $\rm H_2 - H_2$ CIA opacity from \citet{borysow2001, borysow2002} and $\rm H_2-He$ opacity from \citet{borysow1988, borysow1989a, borysow1989b}. The mean molecular weight is calculated based on the composition of the model.

While an overall scaling factor is generally included in high-resolution retrievals \citep{brogi2019, line2021}, previous results from KPIC have found this factor is often poorly constrained from $K$ band data alone \citep[e.g.][]{finnerty2023, finnerty2024}, which for planets with well-constrained radii can be addressed using a restrictive prior on the scale factor \citep[e.g.][]{finnerty2025b, finnerty2025c}. We therefore adopt a log-normal prior with a mean of 1 standard deviation of 0.1 dex for the scale factor.

\section{Results}\label{sec:res}

Parameters, priors, and retrieved values are presented in Table \ref{tab:priors} for the six omitted component cases, and the median and maximum-likelihood spectra obtained from the 6 PC retrievals are plotted in Figure \ref{fig:speccomp}. The full corner plots for the retrievals are included in Appendix \ref{app:corner}. In this section we first discuss the results of the retrievals, followed by the results from cross-correlating the retrieved maximum-likelihood models with the observed time series. 

\begin{figure*}
    \centering
    \includegraphics[width=0.95\linewidth]{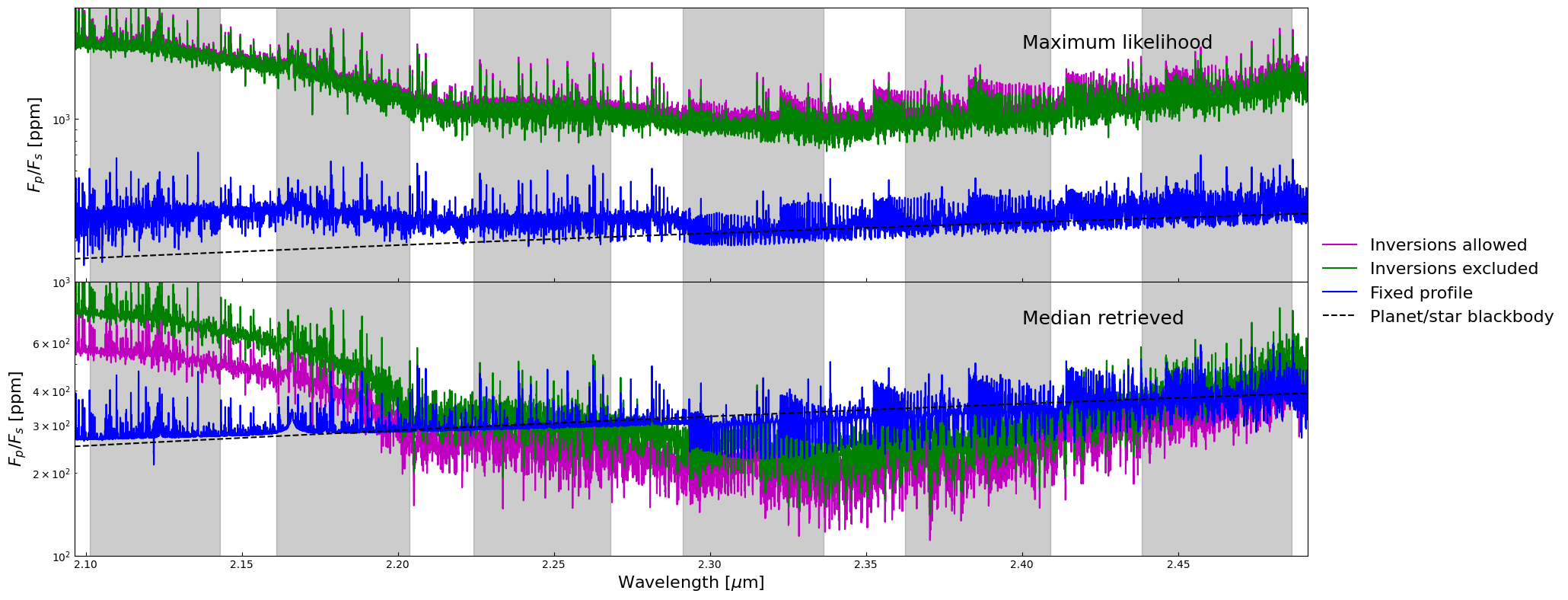}
    \caption{Retrieved maximum-likelihood (top) and median (bottom) spectral models from the retrievals with six omitted components for each of the considered $P-T$ priors. The expected flux ratio is plotted in dashed black, assuming the star and planet are blackbodies at 5500 K and 1500 K, respectively. The fixed profiles cases roughly matches the nominal flux ratio by construction, while the median retrieved free $P-T$ results prefer slightly higher fluxes at short wavelengths and the maximum-likelihood models show significantly stronger fluxes. Such flux offsets are common in short-bandpass HRCCS observations \citep[e.g.][]{finnerty2023, finnerty2025c}. The apparent emission features are a result of absorption features in the stellar model, as all plotted planet models lack a thermal inversion. The spectra show some absorption from CO and CH$_4$, but these features are relatively weak.}
    \label{fig:speccomp}
\end{figure*}

\subsection{Retrieval analysis}

\subsubsection{Significance of retrieval results}

Retrieval analysis does not provide a traditional statistical significance of the planet detection. For HRCCS observations, significance is usually assessed in the \kpvsys\ space, which we discuss in the following subsection. While ideally, retrievals on a dataset without planet features above the noise floor of the data would simply return a flat posterior, in practice a retrieval attempts to find a model that fits \textit{any} features in the data. If planet features are too weak, this may lead to fitting of uncorrected stellar or telluric features, or noise features which slip through the detrending process. Such cases usually result in an obviously non-physical result, particularly if loose priors are used, but this is not guaranteed. In the case of our observations of \hdb, the retrieval analysis returns a posterior that is generally consistent with the expected planetary ephemeris from \citet{pearson2022} and the composition from \citet{sikora2024}, who used data from a different facility with a different analysis technique. The consistency of our results with \citet{sikora2024} is not surprising, given that our analysis provides only weak constraints on the properties of \hdb. Additional observations are needed before we can definitively confirm or refute the \textit{JWST} results through a robust high-resolution detection.  

\subsubsection{Number of Principal Components}

For each of the three treatments of the $P-T$ profile, we performed retrievals omitting 4, 6, and 8 principal components. The posteriors are all generally consistent for a given $P-T$ treatment regardless of the number of omitted PCs. Regardless of the number of omitted components, the \kp\ and \dvsys\ posteriors (see Figures \ref{fig:fixedcorner} and \ref{fig:invcorner}) are bi-modal, with one mode close to the nominal planetary reference frame and the second mode offset by $>20$ \kms. The 4 PC case shows the strongest preference for the offset mode, and the 8 component case the strongest preference for the mode closer to the nominal planet reference frame. This suggests that the offset mode may be the result of incomplete correction of time-varying features when only 4 PCs are omitted. This interpretation is supported by examination of the 2D cross-correlation time series shown in Figure \ref{fig:vtracks}, which shows that omitting only 4 PCs leaves significant cross-correlation features near the stellar/telluric reference frame compared to omitting 6 or 8 PCs. As a result, we omit the 4 component case from subsequent discussion and adopt the 6 component retrieval as our nominal case, and list the retrieved parameters omitting six components in Table \ref{tab:priors}.

\subsubsection{$P-T$ profile treatment}

Whether or not we exclude thermal inversions from the prior range has minimal impact on the retrieval results, with the inversions-permitted case preferring a non-inverted $P-T$ profile at 80\% confidence (80\% of posterior samples have $\log \gamma < 0$) for 6 omitted PCs. The maximum-likelihood abundances listed in Table \ref{tab:priors} are larger in the case where thermal inversions are permitted versus the inversions-excluded retrieval, but this is counteracted by changes in the $P-T$ profile such that the max-likelihood spectra from each retrieval (plotted in Figure \ref{fig:speccomp}) are nearly identical. The median retrieved parameters are in good agreement, as are the median spectra. The continuum level of the median retrieved models are in good agreement with expectations for a 1500 K blackbody, while the maximum-likelihood models have a substantially higher continuum level due to an overall hotter $P-T$ profile. Similar offsets in continuum level have been seen in other $K$-band HRCCS studies \citep[e.g.][]{finnerty2025a, finnerty2025c}. Since observations are continuum-normalized during HRCCS analysis, narrow bandpasses result in an effective loss of continuum information, as HRCCS is sensitive to the strength of observed planetary features relative to the total star+planet continuum level. The results from the fixed $P-T$ profile case are consistent with the free profile cases, and the continuum level is consistent with a 1500 K blackbody by construction.

That the median retrieval prefers a $P-T$ profile roughly consistent with \citet{sikora2024} regardless of whether thermal inversions are permitted suggests that the atmosphere of \hdb\ does not have a strong thermal inversion. Such an inversion would likely produce much stronger planetary spectral features than those preferred by our retrievals, which we expect the retrieval would strongly prefer over the weak absorption features we actually retrieve. In general, the retrieval results are consistent with the results from \textit{JWST} post-eclipse emission spectroscopy reported by \citet{sikora2024}. That work reported a non-inverted $P-T$ profile, strongly detected CH$_4$ absorption features, and made weaker detections of both H$_2$O and CO absorption. Our results are similar, except that the posterior mode consistent with the nominal reference frame of \hdb\ does not prefer significant H$_2$O features. Over the $K$ band, the CH$_4$ opacity is generally greater than the H$_2$O opacity, and if the CH$_4$ abundance is similar to the H$_2$O abundance this could effectively mask H$_2$O features in the Keck/NIRSPEC observations.

\subsection{Cross-correlation results}

\subsubsection{Maximum-likelihood models}

\begin{figure*}
    \centering\includegraphics[width=0.3\linewidth]{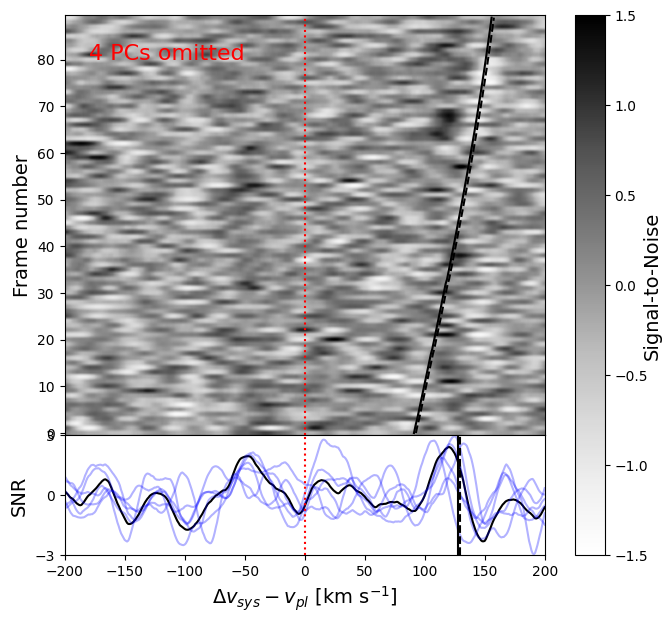}
    \includegraphics[width=0.3\linewidth]{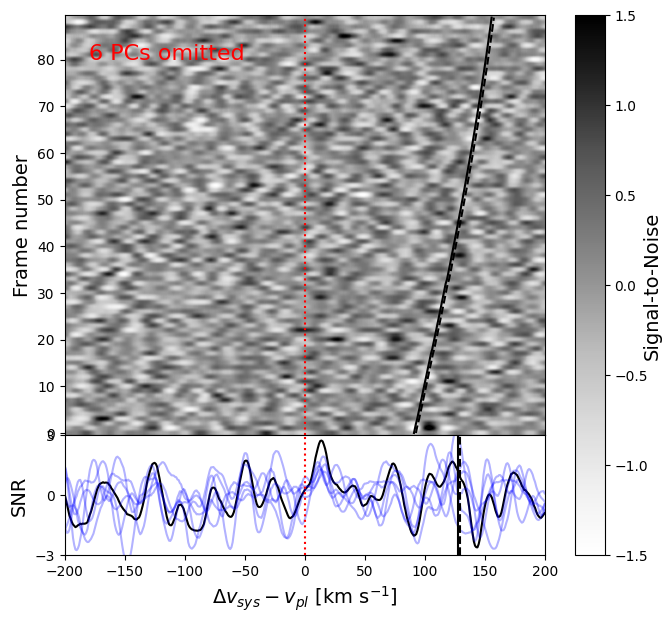}
    \includegraphics[width=0.3\linewidth]{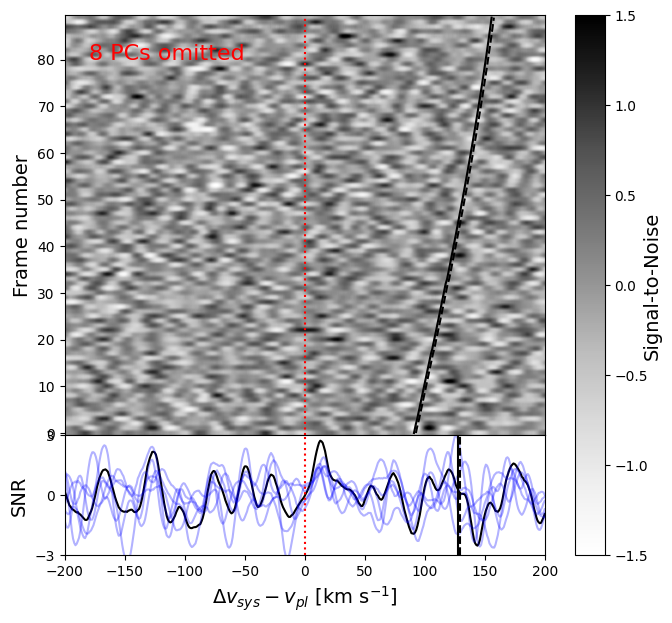}
    \caption{Cross-correlation versus frame number (top) and coadded in the nominal reference frame of \hdb\ (bottom) for the maximum-likelihood inversion-permitted models from the retrievals omitting 4, 6, and 8 PCs from left to right. The nominal planet rest frame is indicated in dashed red, and the stellar and telluric reference frame in solid and dashed black, respectively. Individual orders are shown in the lower panel in blue, and the average of all orders in black. Values are converted to signal-to-noise ratio by dividing by the standard deviation of the region with $|\Delta v_{sys} - v_{pl}| > 50$ \kms. The planet model is reprocessed with the removed PCs by assuming the nominal \kp\ and varying \dvsys. Residuals in the stellar/telluric reference frame are clear in the 4 PC case, suggesting uncorrected non-planetary features are contributing to the cross-correlation, but are much less prominent in the 6 and 8 PC cases. The tentative planet peak is weakly visible after coadding at $\sim10$ \kms, but does not dominate the cross-correlation signal. Note the lower SNR compared with the \kpvsys\ diagrams in Figure \ref{fig:kpvsys} due to differences in the noise estimation.}
    \label{fig:vtracks}
\end{figure*}

\begin{figure*}
    \centering
    \includegraphics[width=0.3\linewidth]{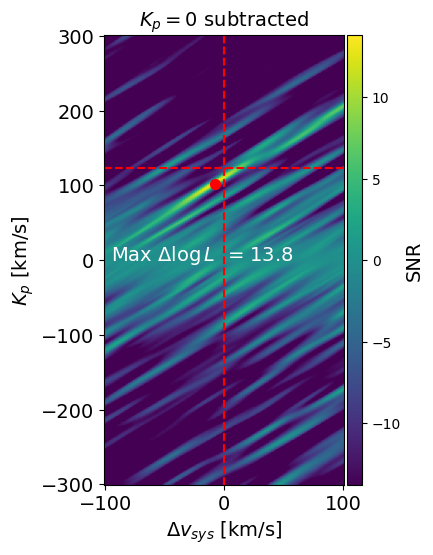}
    \includegraphics[width=0.3\linewidth]{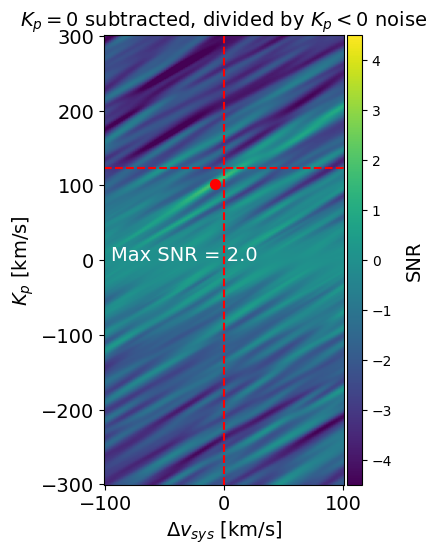}
    \includegraphics[width=0.3\linewidth]{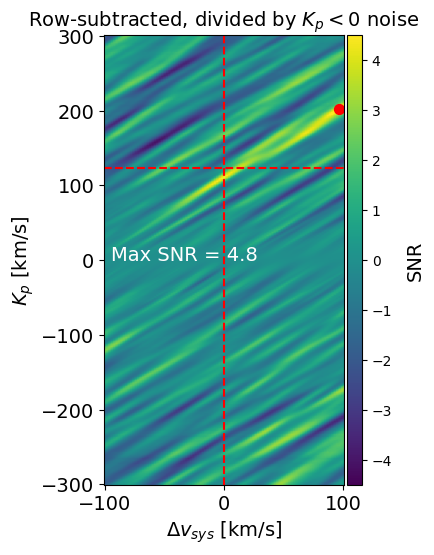}
    \includegraphics[width=0.3\linewidth]{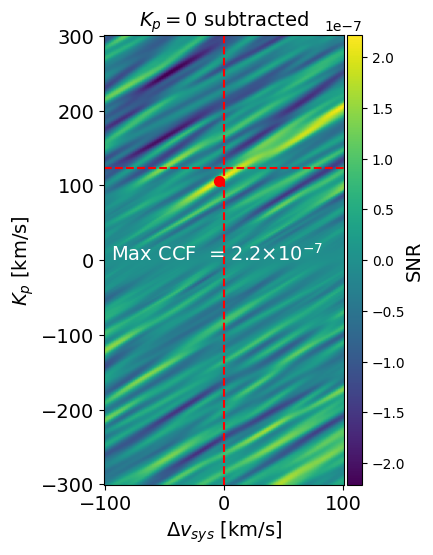}
    \includegraphics[width=0.3\linewidth]{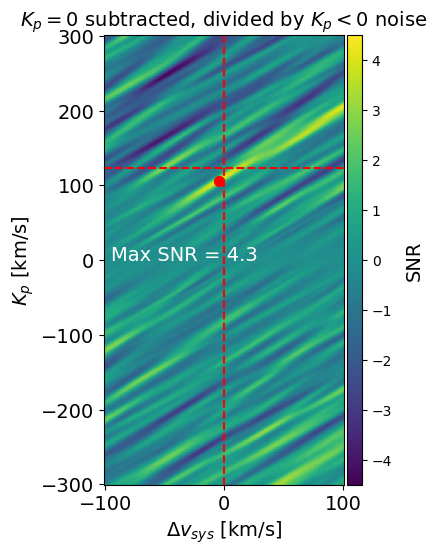}
    \includegraphics[width=0.3\linewidth]{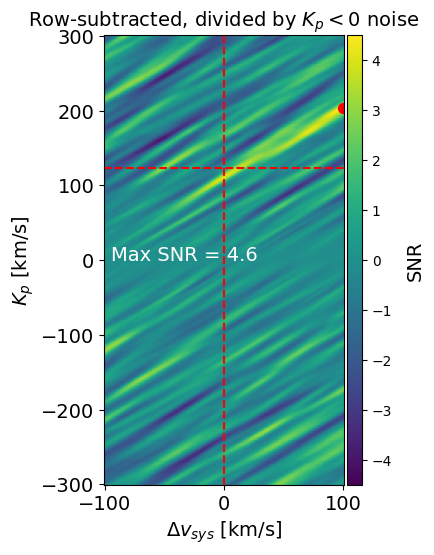}
    \caption{\kpvsys\ maps for the maximum-likelihood model from the 6 PC omitted, non-inverted $P-T$ profile retrieval. The top row shows the maps computed with the \citet{brogi2019} log-likelihood, while the bottom show the cross-correlation coefficient. In both cases, the model is reprocessed at each \kp\ and \dvsys, as is done in the retrievals. The left column shows the raw $\log L$/CCF values after subtracting the median value of the \kp$=0$ row. The center column is the same as the left, but after dividing by the standard deviation of the \kp$<0$ region. The right column shows the maps after median-subtracting each row of constant \kp, then dividing by the standard deviation of the \kp$<0$ region.  The log-likelihood function shows a clear systematic trend to lower values with increasing absolute values of \kp, while both the log-likelihood and the CCF show a smaller variance at smaller absolute values of \kp. In no case do the values in the \kp$<0$ region follow a Gaussian distribution, complicating the interpretation of the SNR, but the treatments similar to previous HRCCS work all suggest a tentative detection with $\rm SNR > 4$.  }
    \label{fig:kpvsys}
\end{figure*}

We assess detection strength using the maximum-likelihood models, which is the best model actually drawn during sampling. In the case of a multi-modal posterior, the median parameters may not correspond to a region of any significant posterior density, making the median model potentially unrepresentative of the planet signal. Figure \ref{fig:vtracks} shows the 2D cross-correlation time series in the nominal reference frame of \hdb\ for 4, 6, and 8 omitted PCs using the maximum-likelihood model from the corresponding free-profile retrieval with thermal inversions permitted. The cross-correlation is converted to a signal-to-noise estimate by dividing the map by the standard deviation of the cross-correlation function in the $|\Delta v_{sys} - v_{pl}| > 50$ \kms\ region. The lower panels show the sum over the time series. Omitting only 4 PCs leaves clear cross-correlation residuals near the stellar/telluric reference frame, and no features of any significance near the nominal reference frame of \hdb. Omitting 6 or 8 PCs largely removes these features from the cross-correlation map. The planet trace is not clear in the 2D cross-correlation function. There is a SNR$\sim$2 peak after summing over the time series, but this is significantly below the SNR$>$4 detection threshold generally used in HRCCS applications, and other features of comparable strength are present at a range of velocities.

Alternatively, we can compute the \kpvsys\ diagram and assess the detection strength based on the strength of potential planet signal compared to the variations far from the nominal planet reference frame. This has been the standard method for assessing HRCCS detection strength since \citet{brogi2012}.  We can compute this map using either the log-likelihood or the the cross-correlation coefficient as a function of varying \kp\ and \dvsys, which we show in Figure \ref{fig:kpvsys}. Figure \ref{fig:kpvsys} includes several different approaches to estimating the noise level in the \kpvsys\ map, which we discuss in detail in Section \ref{sec:disc}. The most common approach taken in HRCCS analysis is shown in the lower middle panel of Figure \ref{fig:kpvsys}. The maximum value of the \kpvsys\ map is offset to higher \kp\ and \dvsys, but the covariance between these parameters is such that this offset is less significant in terms of offset from the nominal planet reference frame. The nominal \kp\ and \dvsys\ are indicated in dashed red in Figure \ref{fig:kpvsys}, and there is a clear peak near the nominal planet reference frame of only marginally lower SNR than the maximum in the entire \kpvsys\ diagram (4.20 compared with 4.62). We consider this a tentative detection of the retrieved maximum-likelihood model, as it falls in the $4 < \rm SNR < 6$ range and there are other correlation and anti-correlation features of comparable strength in the \kpvsys\ map. The maximum-likelihood model from the 8 omitted PC retrieval produces a very similar \kpvsys\ map, indicating that these results are not strongly dependent on the detrending assumptions once more than 4 PCs are removed.

\subsubsection{Molecular models}

\begin{figure*}
    \centering
    \includegraphics[width=0.3\linewidth]{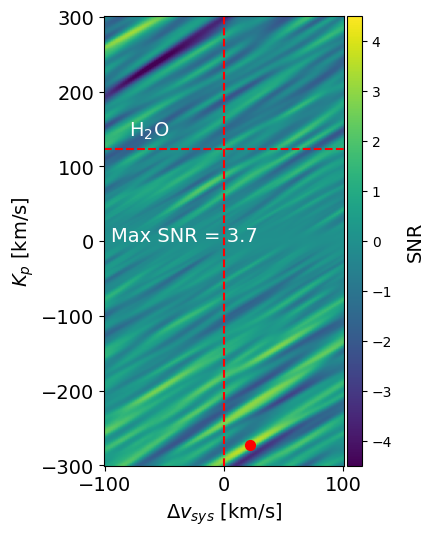}
    \includegraphics[width=0.3\linewidth]{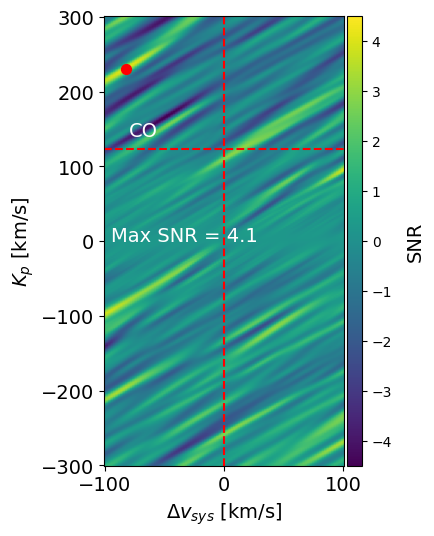}
    \includegraphics[width=0.3\linewidth]{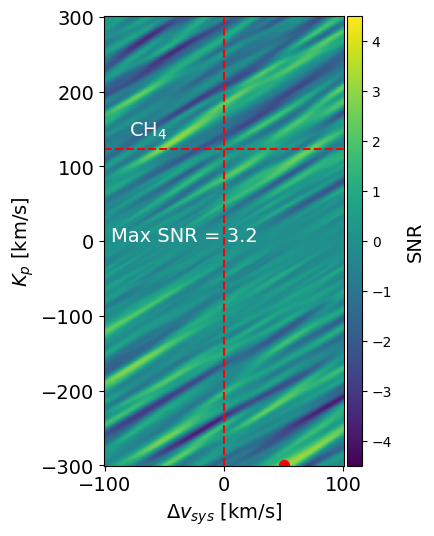}
    \caption{\kpvsys\ maps for individual molecules based on the maximum-likelihood $P-T$ parameters from the 6 PC, no-inversions retrieval. These maps were calculated using the cross-correlation and dividing by the standard deviation of \kp$<0$, consistent with the center bottom panel of Figure \ref{fig:kpvsys}. Consistent with the tentative $\rm SNR\sim4$ detection of the maximum likelihood model, CH$_4$ and CO show non-significant $\rm SNR\sim2$ features consistent with the nominal \kp\ and \dvsys\ of \hdb, indicated by the dashed red lines. H$_2$O does not show significant features in the nominal planetary reference frame. These results are consistent with the retrieved posterior preferring CH$_4$ and CO absorption features while rejecting H$_2$O absorption.}
    \label{fig:kpvsysmols}
\end{figure*}

We assess the detection of H$_2$O, CO, and CH$_4$ individually using models with the same parameters as the maximum-likelihood model from the 6 component, inversions-permitted retrieval, but setting the abundance of one species to $10^{-3}$ and the other two to $10^{-15}$. The resulting \kpvsys\ maps, computed using the CCF and with noise taken from the entire \kp$<0$ region, are shown in Figure \ref{fig:kpvsysmols}. 

No individual molecule model is significantly detected, or even satisfies the $\rm SNR > 4$ criteria to be considered an independent tentative detection. For H$_2$O, this is consistent with the retrieved posterior, which rejects significant H$_2$O features. For CO and CH$_4$, the retrieved posterior suggests we should see these molecules in the \kpvsys\ space, and we do see weak $\rm SNR\sim2$ features coincident with the tentative planet detection in the full maximum-likelihood model, though the maxima of the individual molecule cases are significantly offset from the nominal planet reference frame. 

CH$_4$ has significantly greater overall opacity than H$_2$O or CO throughout most of the $K$-band, but CH$_4$ line features are comparatively weak. As a result, CH$_4$ can act as a pseudo-continuum opacity source, which can be seen in the increased $F_p/F_s$ in the bluest two NIRSPEC orders in Figure \ref{fig:speccomp}.As a result, omitting CH$_4$ in the H$_2$O-only templates changes the baseline planet flux and the relative strengths of H$_2$O lines compared to models including both CH$_4$ and H$_2$O. This could negatively impact the detectability of the single-molecule templates compared with expectations from the all-opacity models.  Given the tentative detection of the all-molecule model, $\rm SNR\sim2-3$ for individual molecules is not surprising, and the molecular \kpvsys\ maps are consistent with our expectations based on the retrieved posteriors..

\section{Discussion}\label{sec:disc}

\subsection{Detection strength estimates}

Estimating the noise level in the \kpvsys\ diagram is non-trivial. As Figure \ref{fig:kpvsys} shows, using the log-likelihood introduces a strong systematic effect with \kp, which requires each row of constant \kp\ be median-subtracted to remove (right column of Figure \ref{fig:kpvsys}). Furthermore, both the log-likelihood and the CCF have lower variance at lower \kp as a result of the PCA reprocessing of the forward model, potentially creating a bias towards higher absolute \kp\ in these maps. This also raises questions about what region should be used to calculate the noise level, as the scale of non-planetary fluctuations (i.e. the noise we are attempting to measure) is itself a function of \kp. Plausible choices include the entire \kp$<0$ region in an attempt to average over the systematic variations, or a range $-K_{p,nominal} \pm\Delta K_p$ over which the systematic effects are negligible. The validity of a particular choice can be assessed by comparing the actual distribution of values in the chosen region with a Gaussian distribution of the same mean and standard deviation, for example through a Kolmogorov-Smirnov (KS) test.

In the case of our observations of \hdb, no choice of noise region we examined satisfied the KS test, which always rejected the noise being Gaussian-distributed with a $p$-value $\lesssim 10^{-9}$. Choosing a small region $-110 < K_p < -130$ \kms\ resulted in $p\sim10^{-9}$, while using the entire \kp$<0$ region resulted in $p\sim10^{-15}$. This non-Gaussian distribution is the case for both the CCF and the log-likelihood maps, and persists regardless of whether or not rows of constant \kp\ are individually median-subtracted. The actual maximum value in the \kpvsys\ map after dividing by the estimated noise is nearly always in the 4-5 range, suggesting that the choice of noise region does not significantly impact the final interpretation of the \kpvsys\ diagram. However, it is worth emphasizing that this ``signal-to-noise" estimate is not equivalent to ``$\sigma$'', which requires that the noise be Gaussian-distributed, and should not be interpreted as such.

Examination of other data sets, including those with very strong planet detections \citep{smith2024, finnerty2025b}, suggests this non-Gaussian off-peak distribution is independent of the particular dataset, and is instead a general feature of this analysis procedure. Examination of the CCF histograms reveals the actual distribution in the \kp$<0$ region is more sharply peaked than the Gaussian of the same mean and variance, with a more extended wing to extreme values. The observed distribution of CCF values is not well-fit by any of a normal, Cauchy, beta, or gamma distribution. 

The extended wings of the CCF distribution provide a possible explanation for some of the anomalous statistical behavior seen in HRCCS analyses. HRCCS detections are generally considered ``tentative'' in the $4 < \rm SNR < 6$ range, as spurious features in the $\rm SNR = 3-4$ range are much more common than would be expected if the noise were Gaussian-distributed and the SNR could be interpreted as a $\sigma$-value. If we could find a well-fitting analytical representation of the off-peak CCF distribution, that probability density function could be used to estimate the chance of obtaining a value as or more extreme than a potential planet feature, a $p$-value, which could then be converted to $\sigma$ for the purposes of comparing false-alarm probabilities. Determining a suitable analytic model for the distribution is beyond the scope of this work. 

\subsection{Injection-recovery tests}
\begin{figure*}
    \centering
    \includegraphics[width=0.35\linewidth]{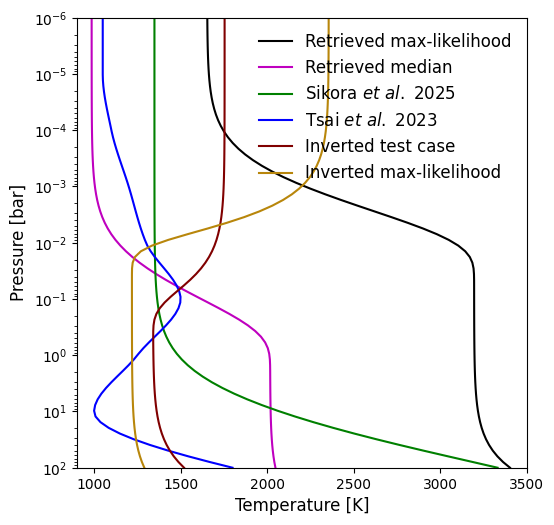}
    \includegraphics[width=0.60\linewidth]{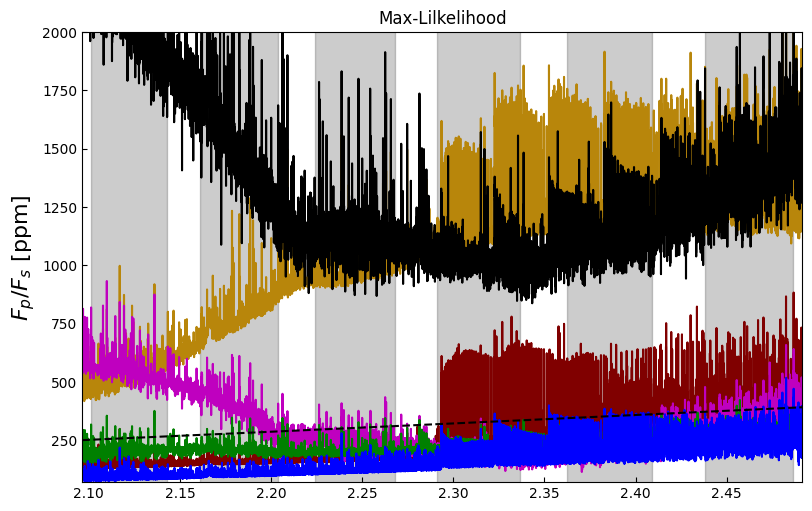}
    \caption{$P-T$ profiles (left) and $F_p/F_s$ for the models used in injection/recovery tests. The absolute temperature of the $P-T$ profile sets the continuum level, which is normalized in HRCCS data processing, while the thermal contrast and abundances jointly set the strength of molecular  features relative to the continuum. Even relatively small thermal inversions ($\sim300$ K) can produce significant emission features, but our retrieval analysis prefers a non-inverted $P-T$ structure.  }
    \label{fig:specplot}
\end{figure*}

\begin{figure*}
    \centering
    \includegraphics[width=0.3\linewidth]{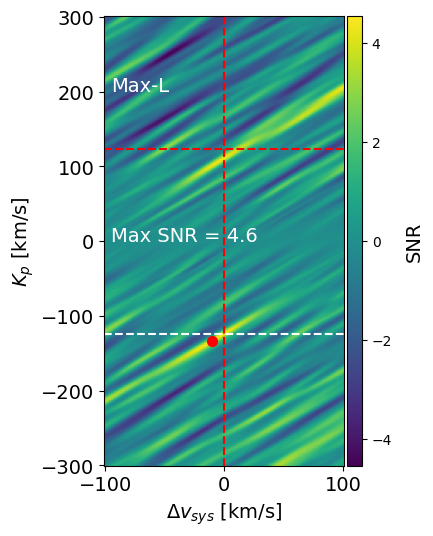}
    \includegraphics[width=0.3\linewidth]{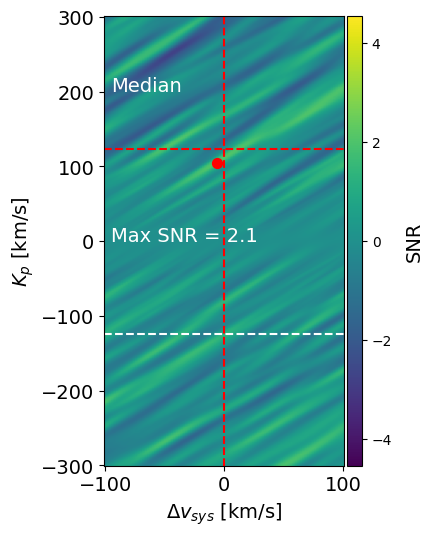}
    \includegraphics[width=0.3\linewidth]{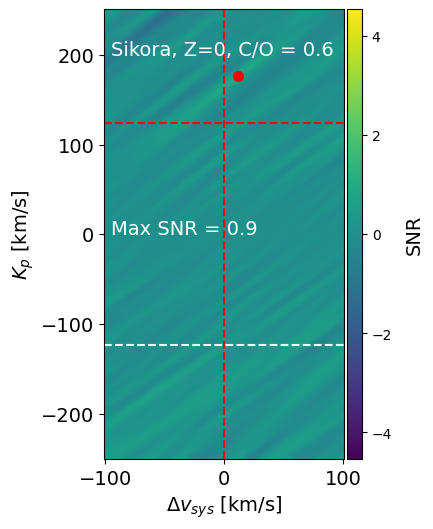}
    \includegraphics[width=0.3\linewidth]{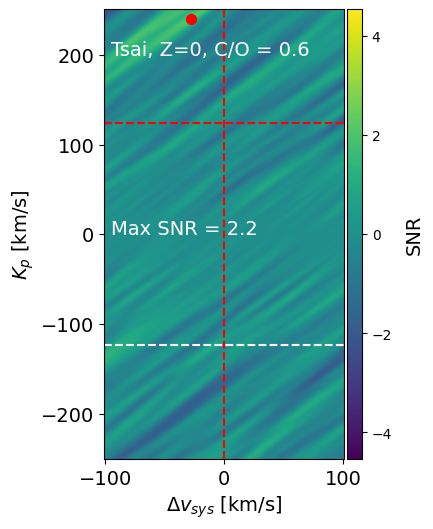}
    \includegraphics[width=0.3\linewidth]{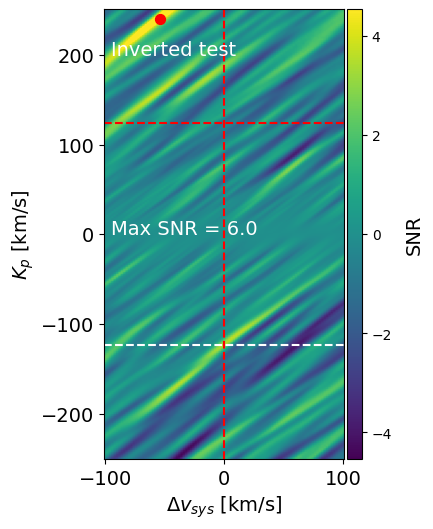}
    \includegraphics[width=0.3\linewidth]{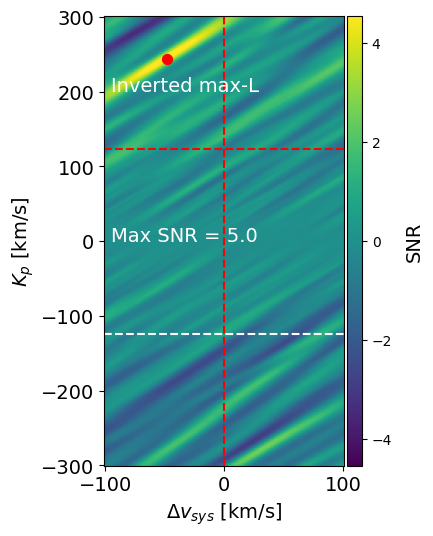}
    \caption{\kpvsys\ diagrams for injection-recovery tests. Top row, from left to right, shows the retrieved maximum-likelihood model, the median retrieved model (both in the 6 component, no inversions case), and a model based on \citet{sikora2024}. Bottom row shows a model based on \citet{tsai2023} at left and an inverted model at right. Model parameters are listed in Table \ref{tab:injmodels}, and plotted in Figure \ref{fig:specplot}. The maps use the same noise level and color scale as the bottom center panel of Figure \ref{fig:kpvsys} to facilitate comparisons. The planet is injected at $-$\kp, indicated in dashed white. The maximum-likelihood and median models are recovered at approximately the same strength as the tentative planet feature they produce, while the features in the \citet{sikora2024} and \citet{tsai2023} models are too weak to produce any significant CCF structure. The inversion test model is recovered at $\rm SNR\sim3.6$, suggesting these data are sensitive to the presence of even relatively weak inversions. 
    }
    \label{fig:injtests}
\end{figure*}

\begin{deluxetable}{ccccc}
    \tablehead{Parameter  & Sikora & Tsai & Inverted test & Inverted max-L }
    \startdata
        $\log \kappa$ & $-2.5$ & - & $-1.5$  & $-1.0$ \\ 
        $\log \gamma$  & $-2.0$ & - & $0.5$  & $1.2$ \\
        $\rm T_{int}$ [K] & 300 & - & 300    & $162$ \\
        $\rm T_{equ}$ [K] & 1600 & - & 1500  & $1430$ \\
        $ K_{\rm p}$  [\kms] & $-123.8$ & $-123.8$ & $-123.8$ & $-123.8$ \\
        $\Delta v_{\rm sys}$ [\kms] & $0$ & $0$ & $0$  & $0$ \\
        $v_{\rm rot}$  [\kms] & 3 & 3 & 3  & $10.4$ \\
        log H$_2$O  &    $-3.7$  & $-3.9$  & $-3.9$   & $-2.4$ \\
        log CO &         $-3.5$  & $-3.3$  & $-3.3$   & $-10.3$ \\
        log CH$_4$  &    $-4.5$  & $-4.5$  & $-7.8$   & $-1.9$ \\
        $\log \rm H_2$ & $-0.13$ & $-0.13$ & $-0.13$  & $-0.2$ \\
        $\log a$ & 0. & 0. & 0. & 0.
    \enddata
    \caption{Parameters for additional used for the injection tests. $P-T$ profiles and spectra are shown in Figure \ref{fig:specplot}. Abundances are based on \texttt{easyCHEM} \citep{lei2024} equilibrium chemistry for the adopted $P-T$ profile. The $P-T$ profile for the model based on \citet{tsai2023} uses a pointwise parameterization. The inverted max-L case is the maximum likelihood inverted model drawn during sampling, while the inverted test case is based on equilibrium chemistry.}
    \label{tab:injmodels}
\end{deluxetable}

The retrieval analysis provides, at best, a tentative detection of \hdb. We assess the plausibility of this potential detection and its implications for the atmosphere of \hdb\ through injection/recovery tests with the retrieved maximum-likelihood and median models and models based on previous results from \citet{sikora2024} and \citet{tsai2023}. The adopted $P-T$ profiles are shown in Figure~\ref{fig:specplot} along with the corresponding $F_p/F_s$. Model parameters are listed in Table \ref{tab:injmodels} for those not based on the retrieved median/max-likelihood. For the abundances, we use \texttt{easyCHEM} to estimate the abundances under chemical equilibrium for each $P-T$ profile, assuming a metallicity of [Fe/H] = 0 and C/O = 0.6, consistent with the results from \citet{sikora2024}. We also consider the maximum-likelihood model drawn during sampling with a thermal inversion, $\log \gamma > 0$. 

We then inject the planet model at nominal strength at $(0,-K_{\rm p})$, which for a planet model well-matched to the data should result in a feature at $(0,-K_{\rm p})$ in the \kpvsys\ map of similar strength to the original detection feature located near $(0,K_{\rm p})$. We compute the \kpvsys\ maps using the cross-correlation, and then convert these maps to SNR using the same noise level as the bottom center panel of Figure \ref{fig:kpvsys}. Figure \ref{fig:injtests} shows the resulting SNR maps on the same color scale to facilitate comparisons.

The retrieved maximum-likelihood model is recovered at a similar signal-to-noise ratio as it is directly detected, suggesting that a planet with a spectrum similar to the maximum-likelihood model is a plausible explanation for the tentative peak near the nominal reference frame of \hdb. A discrepancy between the direct and injected SNR would suggest a mismatch between the maximum-likelihood model and the features being fit, which could happen if the retrieval was attempting to fit uncorrected stellar, telluric, or fringing residuals rather than planetary features. The median retrieved model produces similar results, but the maximum SNR is $<3$. This is consistent with the weaker features in the median retrieved spectrum compared to the maximum-likelihood spectrum (see Figure \ref{fig:specplot}). Furthermore, since the median model is constructed from the posterior, and not actually drawn during sampling, we expect the median model should provide a worse fit to the data than the maximum-likelihood model.

The models based on \citet{sikora2024} and \citet{tsai2023} do not produce any features in the \kpvsys\ map of comparable strength to the retrieved maximum-likelihood model. Figure \ref{fig:specplot} shows that the spectral features in these models are very weak. In the case of the model based on \citet{sikora2024}, the $P-T$ profile has significant thermal contrast, but is nearly isothermal at pressures less than 1 bar, resulting in relatively weak spectral features emerging from the pressure region probed by HRCCS observations. For the \citet{tsai2023} profile, the relatively low temperature sets a low continuum level, and even with a significantly greater $P-T$ contrast compared with the \citet{sikora2024} profile the resulting spectral features remain weak. Given that the median retrieved model is recovered with $\rm SNR < 3$ despite significantly stronger spectral features, the lack of significant features in the \kpvsys\ space with either of the literature-based models is not surprising.

We also consider a model with a thermal inversion of similar strength to the inversion in the \citet{tsai2023} $P-T$ profile, but with the inversion occurring at a lower pressure and a higher deep-atmosphere temperature, and the maximum-likelihood model with an inversion drawn during sampling. These models produce significantly stronger spectral features than any of the models used in injection tests aside from the retrieved maximum-likelihood spectrum. In the injection test, the maximum-likelihood inverted model is recovered at SNR$\sim2$, consistent with the non-inverted models providing a significantly better fit to the data compared with this model. The  constructed test model is marginally better recovered, $3 <  \rm SNR < 4$. While not reaching the $\rm SNR  > 4$ threshold for a tentative detection, these injection tests suggest that if \hdb\ had a strong thermal inversion near periapsis as predicted by theoretical modeling \citep{iro2010, lewis2017, mayorga2021, tsai2023}, these data would not prefer a non-inverted $P-T$ profile. While the injected inverted model is only recovered at $\rm SNR\sim3$, the non-inverted models from the retrieval analysis are similarly detected at $\rm SNR\sim3-5$. As fitting an inverted atmosphere with a non-inverted model would result in a negative cross-correlation, this suggests the maximum-likelihood model is preferred over the inverted model by $6-8\times$ the noise level of the \kpvsys\ map. 

In contrast to the apparent rejection of inverted $P-T$ profiles from the cross-correlation analysis, the retrieved posterior from the inversions-permitted case suggests a $\sim20\%$ probability of an inverted atmosphere ($\log \gamma > 0$). Examination of the posterior samples with $\log \gamma > 0$ finds that these samples are concentrated in a tertiary posterior mode, with \kp$\sim25$ \kms\ and \dvsys$\sim15$ \kms, which can be seen in comparing Figure \ref{fig:invcorner} with Figure \ref{fig:noinvcorner}. This mode corresponds to a strong negative feature in the \kpvsys\ diagrams with the retrieved maximum-likelihood model in Figures \ref{fig:kpvsys} and \ref{fig:injtests}. This large kinematic offset suggests the inverted draws are fitting different features in the data compared with the primary, non-inverted mode. The non-inverted mode is more consistent with the nominal planetary reference frame, and the cross-correlation analysis suggests this mode rejects thermal inversions. The apparent discrepancy between the cross-correlation analysis and the retrievals arises because the cross-correlation analysis is examining a particular signal in the data, whereas the retrieval is comparing all possible signals in the posterior simultaneously.

Both our results and \citet{sikora2024} independently prefer weak absorption features in the atmosphere of \hdb, but additional high-resolution observations are needed to definitively confirm or refute the \textit{JWST} results due to the low detection significance in the existing high-resolution observations. We note that the \citet{sikora2024} posteriors permit inversions at high altitudes ($\rm P < 100$ mbar), above the emission contribution function for low resolution spectroscopy. Compared with \textit{JWST}, high resolution spectroscopy probes significantly higher into the atmosphere, up to the $\sim1$ mbar levels where the \textit{JWST} results permit thermal inversions, and high resolution spectroscopy should therefore be capable of providing strict constraints on the thermal structure of \hdb\ near periapsis. While our tentative results suggest no such inversion is present, we emphasize that additional high-resolution observations are needed to confidently detect \hdb\ and confirm this result independently of the \textit{JWST} observations.

\subsection{Kinematics}

The retrieved maximum-likelihood and median models prefer offsets in both \kp\ and \dvsys\ compared to the nominal ephemeris. Figure \ref{fig:kpvsys} demonstrates that there is a strong covariance between these parameters, as increased \kp\ increases the blueshift from the planet's orbital motion, which can be counteracted by increasing \dvsys\ to introduce a redshift. Assessing the physical significance of the \kp\ and \dvsys\ offsets therefore requires comparing the offset in the retrieved planet velocity at each frame to the nominal ephemeris. For the maximum-likelihood model, the retrieval prefers a redshift increasing from 7 to 19 \kms\ over the course of the observations, while the median values give a redshift from 7 to 15 \kms. This redshift is similar to the 9 \kms\ resolution of NIRSPEC, and the change over the course of the observations is also similar to the detector resolution.

Whether this offset is physically significant is unclear. A 7 \kms\ shift at the start of the observations could be explained by a 15 minute difference between the assumed periapsis time and the actual time of periapsis. This is larger than the 3 minute uncertainty in the periapsis time reported in \citet{pearson2022}, but combined with possible precession and/or relativistic effects on the order of 3--4 minutes \citep{pearson2022} ephemeris errors could explain a significant portion of the retrieved velocity offset. 

Alternatively or additionally, the circulation pattern of the planet could be contributing to the retrieved redshift. At the observed orbital phases, the substellar point of \hdb\ will be rotating out of view, assuming \hdb\ rotates in the same direction that it orbits its host star. If the substellar point dominates the emission spectrum of \hdb, this would result in a net redshift, which could be further compounded by an equatorial jet or day-to-night wind pattern. The actual rotation period of \hdb\ is unknown, which creates significant uncertainty in the scale of these possible effects, but equatorial jets of up to 10 \kms\ have been reported in ultra-hot Jupiters \citep{pai2022}, suggesting circulation patterns could explain at least part of the apparent velocity offset.

The offset from the nominal ephemeris is greatest at the end of the observation. Based on the phase curve from \citet{sikora2024}, we expect the overall planet flux will decrease over the course of the observations as the planet moves past periapsis. Furthermore, the airmass increased over the observations, and conditions began to degrade towards the end of the observations. This combination of factors suggests \hdb\ may not be significantly detected in the later frames, and that the apparent large velocity offset at later times could be due to an absence of detectable planet features which could reject such an offset. 

\subsection{Thermal Structure}

Figure \ref{fig:specplot} shows the median and maximum-likelihood $P-T$ profiles, compared with the fixed $P-T$ profile adopted to roughly match the results from \citet{sikora2024}. The retrieved $P-T$ profile runs hotter than expected, leading to a corresponding increase in $F_p/F_s$. A similar phenomenon has been seen in the retrieval analysis of another marginal HRCCS detection (Finnerty \textit{et al.}, subm.). The HRCCS technique is sensitive to the depths of planet lines relative to the total star+planet continuum level as a result of the continuum being divided out during the data detrending. This can lead to a degeneracy between weak lines at a high planet continuum level and stronger lines at a lower continuum level, as the resulting line depths relative to the host star spectrum can be comparable. 

While the detection of \hdb\ is marginal at best, the presence of a SNR$\sim$4 feature near the expected planet reference frame when correlating with a non-inverted planet model is significant evidence against the presence of a strong thermal inversion in the atmosphere of \hdb, which would be expected to result in a significant negative cross-correlation feature in the planet reference frame when cross-correlating the observations with a non-inverted model atmosphere. Multiple theoretical studies have suggested a thermal inversion forming around the periastron passage of \hdb\ \citep{iro2010, lewis2017, mayorga2021, tsai2023}, but this appears to be at odds with both our results and the \textit{JWST} observations presented in \citet{sikora2024}. This tension between theory and observation is a strong motivator for further observations of \hdb\ with higher spectral resolutions and/or wider bandpasses in order to obtain a clear detection suitable for retrieval analysis.

\subsection{Chemical composition}

Assuming the retrieval is fitting the planetary spectrum, rather than noise, we can make some tentative statements about the composition of \hdb. While the retrieved models are not  strongly detected in cross-correlation, the retrieval analysis still shows weak preferences on the abundances of CO, H$_2$O, and CH$_4$. The posterior mode close to the nominal planet reference frame, which corresponds to the maximum-likelihood model, prefers volume mixing ratios around $10^{-3}$ for CO and CH$_4$, while preferring H$_2$O abundances $\lesssim10^{-3}$. These results are consistent with the single-molecule \kpvsys\ maps in Figure \ref{fig:kpvsysmols}, which show weak features near the nominal planet reference frame for the CO and CH$_4$ models, but no significant features for the H$_2$O template.

These results are  broadly consistent with the \textit{JWST} G395H results from \citet{sikora2024}, who reported a $4-10\sigma$ detection of CH$_4$, $4-5\sigma$ detection of H$_2$O, and $\sim4\sigma$ detection of CO from secondary eclipse spectroscopy, with significant changes in detection strength based on orbital phase. Compared with the $K$-band, the $2.8-5.2\mu\rm m$ G395H bandpass has significantly stronger CH$_4$ opacity features, which may explain the stronger CH$_4$ detection in the \textit{JWST} data. In the case of H$_2$O, the $K$-band has strong opacity features, but the continuum CH$_4$ opacity is significantly greater than the H$_2$O opacity through most of the band. As a result, at equal abundances of CH$_4$ and H$_2$O, the H$_2$O features will be difficult to distinguish, which may explain why the NIRSPEC data do not detect H$_2$O while the \textit{JWST} observations do.

Abundance estimation in \hdb\ is complicated by the possibility of significant vertical abundance gradients. While \hdb\ does not appear to develop a strong thermal inversion near periastron, the possible presence of CO features does suggest changes in atmospheric chemistry with orbital phase, as at apoastron the dominant carbon species is expected to be CH$_4$. The rapidly changing stellar flux experienced by \hdb\ and the apparent discrepancy between theoretical predictions and the observed spectrum suggests a free retrieval approach which incorporates dissociation \citep[e.g.][]{finnerty2025b} may be required for accurate abundance constraints once higher-quality data are available. 

\section{Summary and Conclusions}

We present post-eclipse $K$ band observations of the highly eccentric planet \hdb\ from Keck II/NIRSPEC. Retrieval and cross-correlation analysis suggest a tentative detection of \hdb\ with an absorption-dominated spectrum, without a strong near-periastron thermal inversion as predicted by GCMs. Our retrieval analysis suggests the presence of CH$_4$ and CO absorption features, while H$_2$O is weakly disfavored. These \textit{tentative} results are consistent with recent results from \textit{JWST} observations \citep{sikora2024}, which strongly disfavored a thermal inversion and found evidence for absorption features of CH$_4$, H$_2$O, and CO. Additional observations at high spectral resolution are needed to make a robust detection of \hdb\ and independently confirm the \textit{JWST} results.

The long period of \hdb\ poses significant challenges for ground-based observations. Until 2030, only 2026-03-11 (UT) and 2028-12-07 (UT) periastron passages identified in \citet{pearson2022} will be accessible to observatories on Mauna Kea. The latter event offers good post-eclipse phase coverage, and Keck/HISPEC \citep{hispec} is expected to be available during that time, presenting an opportunity to obtain high-resolution observations of \hdb\ with significantly improved stability, wavelength coverage, and spectral resolution compared with the data presented here. 

\section{Acknowledgments}

We thank the anonymous referee whose detailed comments improved the quality of this paper. L.F. was a member of UAW local 4811 while this research was carried out, and is a member of UM-PRO. Some of the research described in this publication was carried out in part at the Jet Propulsion Laboratory, California Institute of Technology, under a contract with the National Aeronautics and Space Administration. This research has made use of the NASA Exoplanet Archive \citep{10.26133/nea12} and the Exoplanet Follow-up Observing Program \citep{10.26134/exofop5}, which are operated by the California Institute of Technology, under contract with the National Aeronautics and Space Administration under the Exoplanet Exploration Program.

This work used computational and storage services associated with the Hoffman2 Shared Cluster provided by UCLA Institute for Digital Research and Education’s Research Technology Group. The contributed Hoffman2 computing node used for this work was supported by the Heising-Simons Foundation grant \#2020-1821.

The data presented herein were obtained at the W. M. Keck Observatory, which is operated as a scientific partnership among the California Institute of Technology, the University of California and the National Aeronautics and Space Administration. The Observatory was made possible by the generous financial support of the W. M. Keck Foundation.  The authors wish to recognize and acknowledge the very significant cultural role and reverence that the summit of Maunakea has always had within the indigenous Hawaiian community.  We are most fortunate to have the opportunity to conduct observations from this mountain.

\facility{Keck:II (NIRSEPC)}

\appendix
\section{Corner Plots}\label{app:corner}
\begin{figure}
    \centering
    \includegraphics[width=1.0\linewidth]{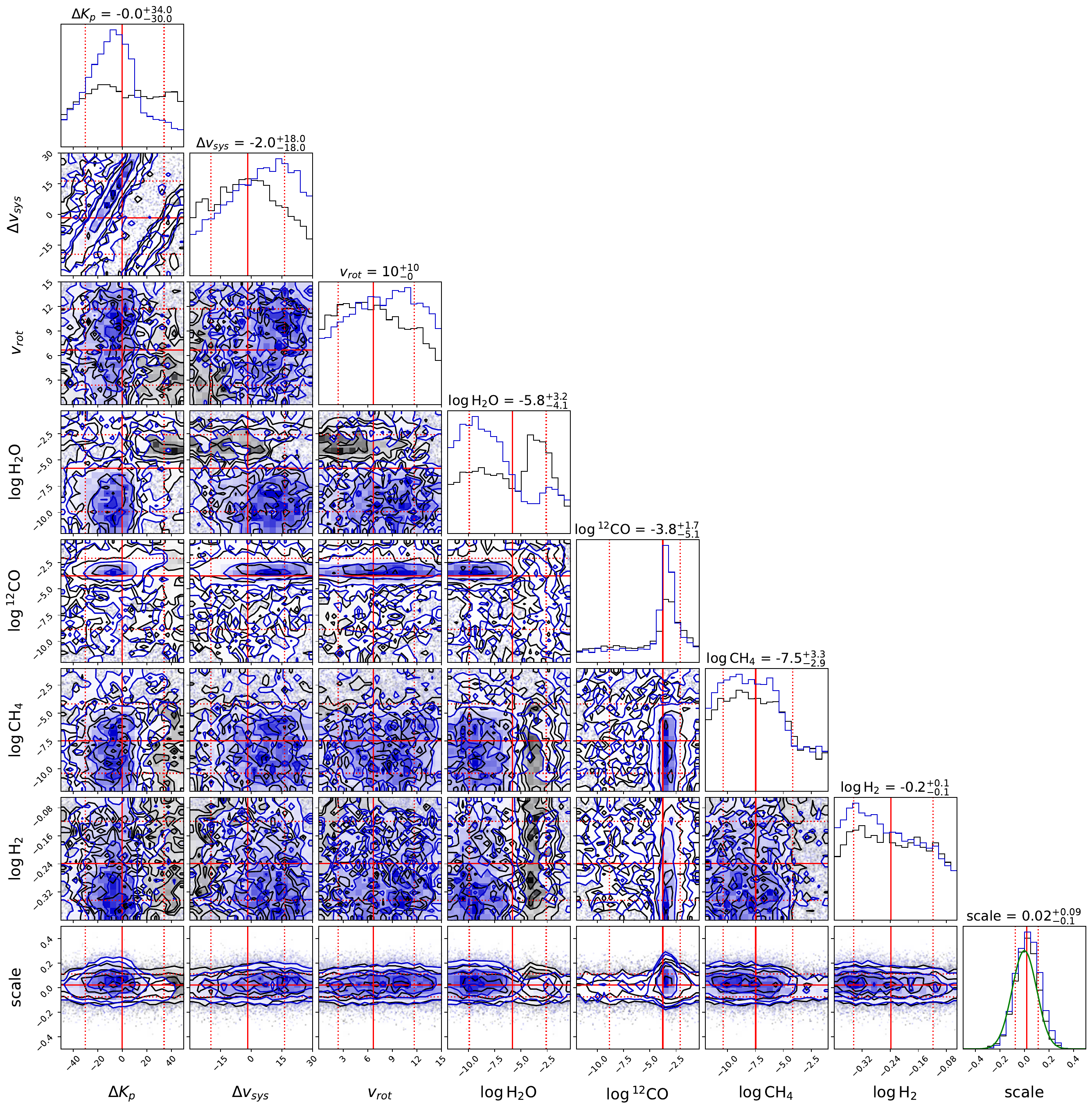}
    \caption{Corner plot for the retrieval with the fixed $P-T$ profile. The black posterior shows the 6 and blue shows 8 PC. Confidence intervals are given for the 6 PC posterior. The retrieved posteriors span the full prior range in both cases. }
    \label{fig:fixedcorner}
\end{figure}

\begin{figure}
    \centering
    \includegraphics[width=1.0\linewidth]{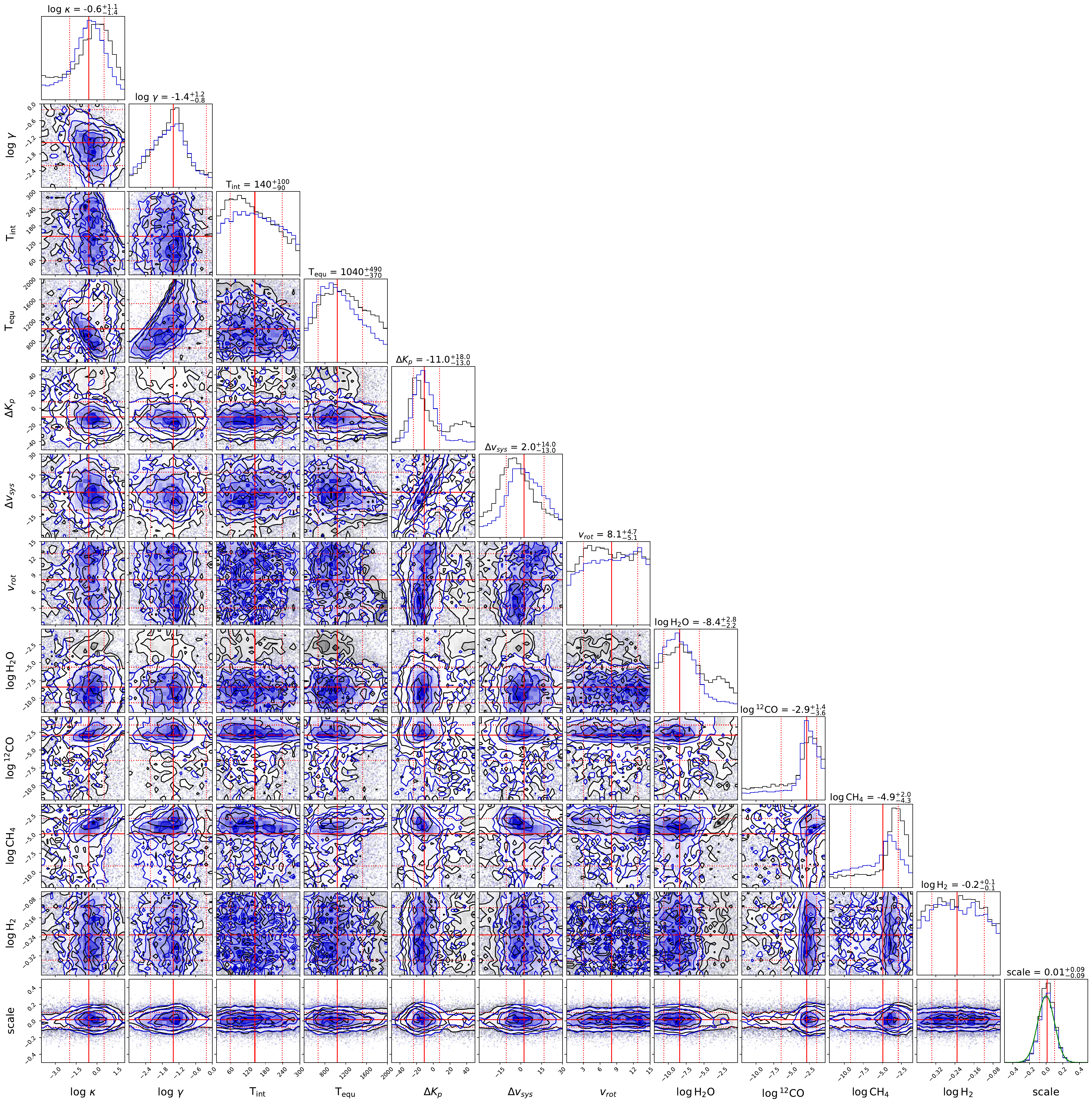}
    \caption{Corner plot for the retrieval with the \citet{guillot2010} $P-T$ profile. The black posterior shows the 6 and blue shows 8 PC. Confidence intervals are given for the 6 PC posterior. As with the fixed-profile case, the 6 and 8 component posteriors are generally consistent with each other. Most parameters are unconstrained. There is clear multimodality and covariance in the \kp\ and \dvsys\ parameters, but one of the modes in the 6 PC case is consistent with the expected reference frame of the planet and the dominant mode of the 6 PC retrieval when inversions are not permitted by prior (see below).}
    \label{fig:invcorner}
\end{figure}

\begin{figure}
    \centering
    \includegraphics[width=1.0\linewidth]{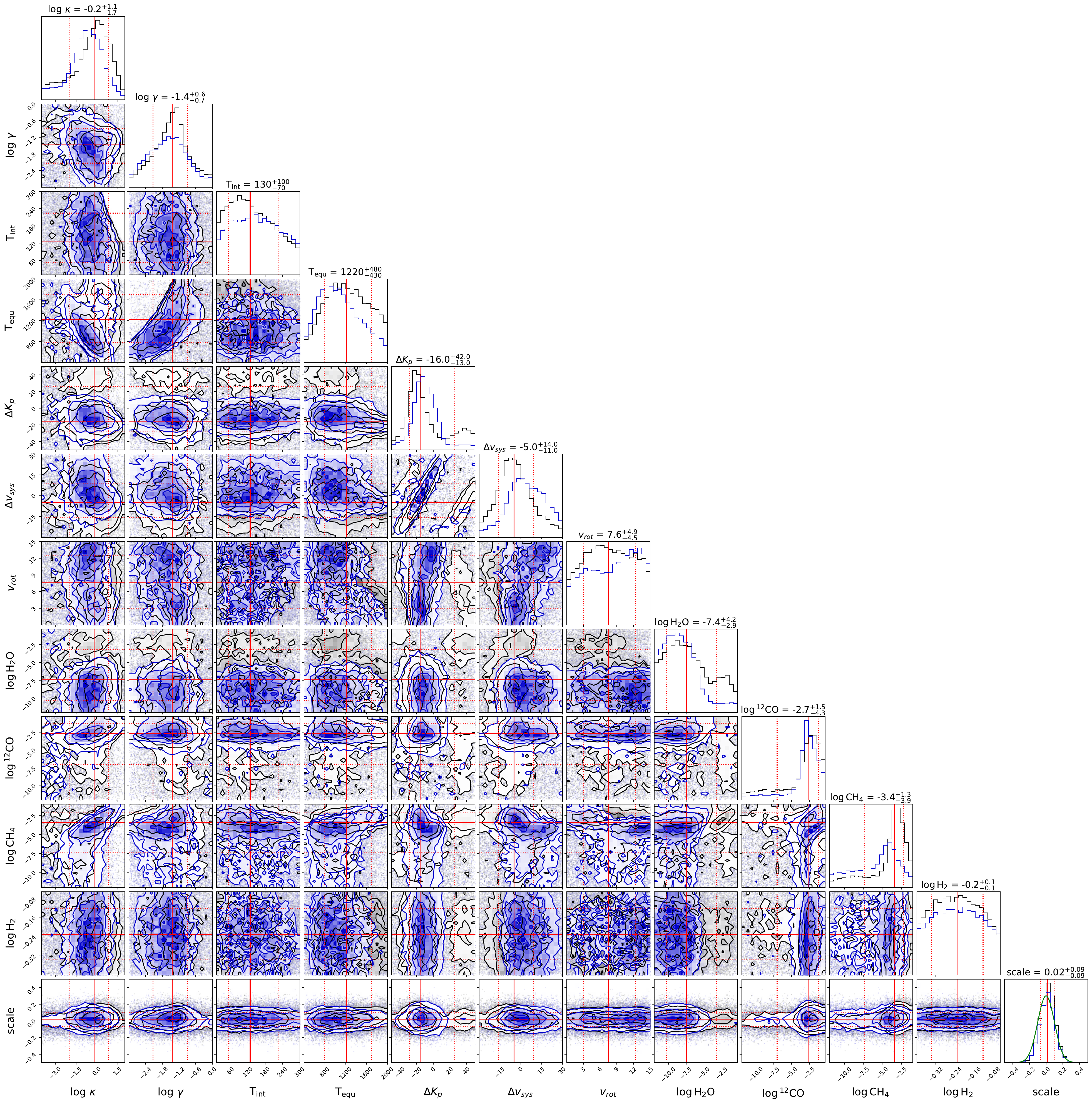}
    \caption{Corner plot for the retrieval with the $P-T$ profile excluding inversions. The black posterior shows the 6 and blue shows 8 PC. Confidence intervals are given for the 6 PC posterior. There are significant discrepancies between the three retrieved posteriors, and the 8 PC case is almost entirely unconstrained. Omitting 6 PCs results in a generally constrained posterior and velocity parameters consistent with the nominal reference frame of \hdb. }
    \label{fig:noinvcorner}
\end{figure}

Figure \ref{fig:fixedcorner} presents the corner plot for the retrievals assuming a fixed $P-T$ profile based on \citet{sikora2024}. Figure \ref{fig:invcorner} presents the corner plot from the retrievals with a loose prior on the $P-T$ profile which permits thermal inversion. Figure \ref{fig:noinvcorner} presents the corner plot from the retrievals with a freely fit $P-T$ profile, excluding inversions as a prior. In each corner plot the 6 and 8 omitted principal component runs are shown in black and blue, respectively. The printed median and $\pm34\%$ confidence intervals are for the six component case. The posteriors in the 6 and 8 component cases generally agree for all $P-T$ approaches, although the 8 PC case is generally more poorly constrained.

\clearpage

\bibliography{exoplanetbib}{}

\begin{thebibliography}{}
\expandafter\ifx\csname natexlab\endcsname\relax\def\natexlab#1{#1}\fi
\providecommand{\url}[1]{\href{#1}{#1}}
\providecommand{\dodoi}[1]{doi:~\href{http://doi.org/#1}{\nolinkurl{#1}}}
\providecommand{\doeprint}[1]{\href{http://ascl.net/#1}{\nolinkurl{http://ascl.net/#1}}}
\providecommand{\doarXiv}[1]{\href{https://arxiv.org/abs/#1}{\nolinkurl{https://arxiv.org/abs/#1}}}

\bibitem[{{Birkby}(2018)}]{birkby2018}
{Birkby}, J.~L. 2018, arXiv e-prints, arXiv:1806.04617, \dodoi{10.48550/arXiv.1806.04617}

\bibitem[{{Bonomo} {et~al.}(2017){Bonomo}, {Desidera}, {Benatti}, {Borsa}, {Crespi}, {Damasso}, {Lanza}, {Sozzetti}, {Lodato}, {Marzari}, {Boccato}, {Claudi}, {Cosentino}, {Covino}, {Gratton}, {Maggio}, {Micela}, {Molinari}, {Pagano}, {Piotto}, {Poretti}, {Smareglia}, {Affer}, {Biazzo}, {Bignamini}, {Esposito}, {Giacobbe}, {H{\'e}brard}, {Malavolta}, {Maldonado}, {Mancini}, {Martinez Fiorenzano}, {Masiero}, {Nascimbeni}, {Pedani}, {Rainer}, \& {Scandariato}}]{bonomo2017}
{Bonomo}, A.~S., {Desidera}, S., {Benatti}, S., {et~al.} 2017, \aap, 602, A107, \dodoi{10.1051/0004-6361/201629882}

\bibitem[{{Borysow}(2002)}]{borysow2002}
{Borysow}, A. 2002, \aap, 390, 779, \dodoi{10.1051/0004-6361:20020555}

\bibitem[{{Borysow} \& {Frommhold}(1989)}]{borysow1989b}
{Borysow}, A., \& {Frommhold}, L. 1989, \apj, 341, 549, \dodoi{10.1086/167515}

\bibitem[{{Borysow} {et~al.}(1989){Borysow}, {Frommhold}, \& {Moraldi}}]{borysow1989a}
{Borysow}, A., {Frommhold}, L., \& {Moraldi}, M. 1989, \apj, 336, 495, \dodoi{10.1086/167027}

\bibitem[{{Borysow} {et~al.}(2001){Borysow}, {Jorgensen}, \& {Fu}}]{borysow2001}
{Borysow}, A., {Jorgensen}, U.~G., \& {Fu}, Y. 2001, \jqsrt, 68, 235, \dodoi{10.1016/S0022-4073(00)00023-6}

\bibitem[{{Borysow} {et~al.}(1988){Borysow}, {Frommhold}, \& {Birnbaum}}]{borysow1988}
{Borysow}, J., {Frommhold}, L., \& {Birnbaum}, G. 1988, \apj, 326, 509, \dodoi{10.1086/166112}

\bibitem[{{Brogi} \& {Line}(2019)}]{brogi2019}
{Brogi}, M., \& {Line}, M.~R. 2019, \aj, 157, 114, \dodoi{10.3847/1538-3881/aaffd3}

\bibitem[{{Brogi} {et~al.}(2012){Brogi}, {Snellen}, {de Kok}, {Albrecht}, {Birkby}, \& {de Mooij}}]{brogi2012}
{Brogi}, M., {Snellen}, I. A.~G., {de Kok}, R.~J., {et~al.} 2012, \nat, 486, 502, \dodoi{10.1038/nature11161}

\bibitem[{{Buchner} {et~al.}(2014){Buchner}, {Georgakakis}, {Nandra}, {Hsu}, {Rangel}, {Brightman}, {Merloni}, {Salvato}, {Donley}, \& {Kocevski}}]{buchner2014}
{Buchner}, J., {Georgakakis}, A., {Nandra}, K., {et~al.} 2014, \aap, 564, A125, \dodoi{10.1051/0004-6361/201322971}

\bibitem[{{Carvalho} \& {Johns-Krull}(2023)}]{carvalho2023}
{Carvalho}, A., \& {Johns-Krull}, C.~M. 2023, Research Notes of the American Astronomical Society, 7, 91, \dodoi{10.3847/2515-5172/acd37e}

\bibitem[{{Cutri} {et~al.}(2003){Cutri}, {Skrutskie}, {van Dyk}, {Beichman}, {Carpenter}, {Chester}, {Cambresy}, {Evans}, {Fowler}, {Gizis}, {Howard}, {Huchra}, {Jarrett}, {Kopan}, {Kirkpatrick}, {Light}, {Marsh}, {McCallon}, {Schneider}, {Stiening}, {Sykes}, {Weinberg}, {Wheaton}, {Wheelock}, \& {Zacarias}}]{cutri2003}
{Cutri}, R.~M., {Skrutskie}, M.~F., {van Dyk}, S., {et~al.} 2003, {2MASS All Sky Catalog of point sources.}

\bibitem[{{de Wit} {et~al.}(2016){de Wit}, {Lewis}, {Langton}, {Laughlin}, {Deming}, {Batygin}, \& {Fortney}}]{dewit2016}
{de Wit}, J., {Lewis}, N.~K., {Langton}, J., {et~al.} 2016, \apjl, 820, L33, \dodoi{10.3847/2041-8205/820/2/L33}

\bibitem[{{Delorme} {et~al.}(2021){Delorme}, {Jovanovic}, {Echeverri}, {Mawet}, {Kent Wallace}, {Bartos}, {Cetre}, {Wizinowich}, {Ragland}, {Lilley}, {Wetherell}, {Doppmann}, {Wang}, {Morris}, {Ruffio}, {Martin}, {Fitzgerald}, {Ruane}, {Schofield}, {Suominen}, {Calvin}, {Wang}, {Magnone}, {Johnson}, {Sohn}, {L{\'o}pez}, {Bond}, {Pezzato}, {Sayson}, {Chun}, \& {Skemer}}]{kpic}
{Delorme}, J.-R., {Jovanovic}, N., {Echeverri}, D., {et~al.} 2021, Journal of Astronomical Telescopes, Instruments, and Systems, 7, 035006, \dodoi{10.1117/1.JATIS.7.3.035006}

\bibitem[{{Ehrenreich} {et~al.}(2020){Ehrenreich}, {Lovis}, {Allart}, {Zapatero Osorio}, {Pepe}, {Cristiani}, {Rebolo}, {Santos}, {Borsa}, {Demangeon}, {Dumusque}, {Gonz{\'a}lez Hern{\'a}ndez}, {Casasayas-Barris}, {S{\'e}gransan}, {Sousa}, {Abreu}, {Adibekyan}, {Affolter}, {Allende Prieto}, {Alibert}, {Aliverti}, {Alves}, {Amate}, {Avila}, {Baldini}, {Bandy}, {Benz}, {Bianco}, {Bolmont}, {Bouchy}, {Bourrier}, {Broeg}, {Cabral}, {Calderone}, {Pall{\'e}}, {Cegla}, {Cirami}, {Coelho}, {Conconi}, {Coretti}, {Cumani}, {Cupani}, {Dekker}, {Delabre}, {Deiries}, {D'Odorico}, {Di Marcantonio}, {Figueira}, {Fragoso}, {Genolet}, {Genoni}, {G{\'e}nova Santos}, {Hara}, {Hughes}, {Iwert}, {Kerber}, {Knudstrup}, {Landoni}, {Lavie}, {Lizon}, {Lendl}, {Lo Curto}, {Maire}, {Manescau}, {Martins}, {M{\'e}gevand}, {Mehner}, {Micela}, {Modigliani}, {Molaro}, {Monteiro}, {Monteiro}, {Moschetti}, {M{\"u}ller}, {Nunes}, {Oggioni}, {Oliveira}, {Pariani}, {Pasquini}, {Poretti}, {Rasilla}, {Redaelli}, {Riva}, {Santana Tschudi},
  {Santin}, {Santos}, {Segovia Milla}, {Seidel}, {Sosnowska}, {Sozzetti}, {Span{\`o}}, {Su{\'a}rez Mascare{\~n}o}, {Tabernero}, {Tenegi}, {Udry}, {Zanutta}, \& {Zerbi}}]{ehrenreich2020}
{Ehrenreich}, D., {Lovis}, C., {Allart}, R., {et~al.} 2020, \nat, 580, 597, \dodoi{10.1038/s41586-020-2107-1}

\bibitem[{{Feroz} \& {Hobson}(2008)}]{feroz2008}
{Feroz}, F., \& {Hobson}, M.~P. 2008, \mnras, 384, 449, \dodoi{10.1111/j.1365-2966.2007.12353.x}

\bibitem[{{Feroz} {et~al.}(2009){Feroz}, {Hobson}, \& {Bridges}}]{feroz2009}
{Feroz}, F., {Hobson}, M.~P., \& {Bridges}, M. 2009, \mnras, 398, 1601, \dodoi{10.1111/j.1365-2966.2009.14548.x}

\bibitem[{{Feroz} {et~al.}(2019){Feroz}, {Hobson}, {Cameron}, \& {Pettitt}}]{feroz2019}
{Feroz}, F., {Hobson}, M.~P., {Cameron}, E., \& {Pettitt}, A.~N. 2019, The Open Journal of Astrophysics, 2, 10, \dodoi{10.21105/astro.1306.2144}

\bibitem[{{Finnerty} {et~al.}(2023){Finnerty}, {Schofield}, {Sappey}, {Xuan}, {Ruffio}, {Wang}, {Delorme}, {Blake}, {Buzard}, {Fitzgerald}, {Baker}, {Bartos}, {Bond}, {Calvin}, {Cetre}, {Doppmann}, {Echeverri}, {Jovanovic}, {Liberman}, {L{\'o}pez}, {Martin}, {Mawet}, {Morris}, {Pezzato}, {Phillips}, {Ragland}, {Skemer}, {Venenciano}, {Wallace}, {Wallack}, {Wang}, \& {Wizinowich}}]{finnerty2023}
{Finnerty}, L., {Schofield}, T., {Sappey}, B., {et~al.} 2023, \aj, 166, 31, \dodoi{10.3847/1538-3881/acda91}

\bibitem[{{Finnerty} {et~al.}(2024){Finnerty}, {Xuan}, {Xin}, {Liberman}, {Schofield}, {Fitzgerald}, {Agrawal}, {Baker}, {Bartos}, {Blake}, {Calvin}, {Cetre}, {Delorme}, {Doppmann}, {Echeverri}, {Hsu}, {Jovanovic}, {L{\'o}pez}, {Martin}, {Mawet}, {Morris}, {Pezzato}, {Ruffio}, {Sappey}, {Skemer}, {Venenciano}, {Wallace}, {Wallack}, {Wang}, \& {Wang}}]{finnerty2024}
{Finnerty}, L., {Xuan}, J.~W., {Xin}, Y., {et~al.} 2024, \aj, 167, 43, \dodoi{10.3847/1538-3881/ad1180}

\bibitem[{{Finnerty} {et~al.}(2025{\natexlab{a}}){Finnerty}, {Xin}, {Xuan}, {Inglis}, {Fitzgerald}, {Agrawal}, {Baker}, {Bartos}, {Blake}, {Calvin}, {Cetre}, {Delorme}, {Doppmann}, {Echeverri}, {Horstman}, {Hsu}, {Jovanovic}, {Liberman}, {L{\'o}pez}, {Martin}, {Mawet}, {Morris}, {Pezzato-Rovner}, {Ruffio}, {Sappey}, {Schofield}, {Skemer}, {Venenciano}, {Wallace}, {Wallack}, {Wang}, \& {Wang}}]{finnerty2025a}
{Finnerty}, L., {Xin}, Y., {Xuan}, J.~W., {et~al.} 2025{\natexlab{a}}, \aj, 169, 94, \dodoi{10.3847/1538-3881/ada1d9}

\bibitem[{{Finnerty} {et~al.}(2025{\natexlab{b}}){Finnerty}, {Xin}, {Xuan}, {Inglis}, {Fitzgerald}, {Agrawal}, {Baker}, {Bartos}, {Blake}, {Calvin}, {Cetre}, {Delorme}, {Doppmann}, {Echeverri}, {Horstman}, {Hsu}, {Jovanovic}, {Liberman}, {L{\'o}pez}, {Martin}, {Mawet}, {Morris}, {Pezzato}, {Ruffio}, {Sappey}, {Schofield}, {Skemer}, {Venenciano}, {Wallace}, {Wallack}, {Wang}, \& {Wang}}]{finnerty2025b}
---. 2025{\natexlab{b}}, \aj, 169, 333, \dodoi{10.3847/1538-3881/adce02}

\bibitem[{{Finnerty} {et~al.}(2025{\natexlab{c}}){Finnerty}, {Inglis}, {Fitzgerald}, {Echeverri}, {Jovanovic}, {Mawet}, {Blake}, {Baker}, {Bartos}, {Calvin}, {Cetre}, {Delorme}, {Doppmann}, {Horstman}, {Hsu}, {Liberman}, {L{\'o}pez}, {Morris}, {Pezzato-Rovner}, {Ruffio}, {Sappey}, {Schofield}, {Skemer}, {Wallace}, {Wallack}, {Wang}, {Wang}, {Xin}, \& {Xuan}}]{finnerty2025c}
{Finnerty}, L., {Inglis}, J., {Fitzgerald}, M.~P., {et~al.} 2025{\natexlab{c}}, arXiv e-prints, arXiv:2508.09448, \dodoi{10.48550/arXiv.2508.09448}

\bibitem[{{Gaia Collaboration}(2020)}]{gaiaedr3}
{Gaia Collaboration}. 2020, VizieR Online Data Catalog, I/350

\bibitem[{{Gordon} {et~al.}(2022){Gordon}, {Rothman}, {Hargreaves}, {Hashemi}, {Karlovets}, {Skinner}, {Conway}, {Hill}, {Kochanov}, {Tan}, {Wcis{\l}o}, {Finenko}, {Nelson}, {Bernath}, {Birk}, {Boudon}, {Campargue}, {Chance}, {Coustenis}, {Drouin}, {Flaud}, {Gamache}, {Hodges}, {Jacquemart}, {Mlawer}, {Nikitin}, {Perevalov}, {Rotger}, {Tennyson}, {Toon}, {Tran}, {Tyuterev}, {Adkins}, {Baker}, {Barbe}, {Can{\`e}}, {Cs{\'a}sz{\'a}r}, {Dudaryonok}, {Egorov}, {Fleisher}, {Fleurbaey}, {Foltynowicz}, {Furtenbacher}, {Harrison}, {Hartmann}, {Horneman}, {Huang}, {Karman}, {Karns}, {Kassi}, {Kleiner}, {Kofman}, {Kwabia-Tchana}, {Lavrentieva}, {Lee}, {Long}, {Lukashevskaya}, {Lyulin}, {Makhnev}, {Matt}, {Massie}, {Melosso}, {Mikhailenko}, {Mondelain}, {M{\"u}ller}, {Naumenko}, {Perrin}, {Polyansky}, {Raddaoui}, {Raston}, {Reed}, {Rey}, {Richard}, {T{\'o}bi{\'a}s}, {Sadiek}, {Schwenke}, {Starikova}, {Sung}, {Tamassia}, {Tashkun}, {Vander Auwera}, {Vasilenko}, {Vigasin}, {Villanueva}, {Vispoel}, {Wagner}, {Yachmenev}, \&
  {Yurchenko}}]{hitemp2020}
{Gordon}, I.~E., {Rothman}, L.~S., {Hargreaves}, R.~J., {et~al.} 2022, \jqsrt, 277, 107949, \dodoi{10.1016/j.jqsrt.2021.107949}

\bibitem[{{Guillot}(2010)}]{guillot2010}
{Guillot}, T. 2010, \aap, 520, A27, \dodoi{10.1051/0004-6361/200913396}

\bibitem[{{Hargreaves} {et~al.}(2020){Hargreaves}, {Gordon}, {Rey}, {Nikitin}, {Tyuterev}, {Kochanov}, \& {Rothman}}]{hargreaves2020}
{Hargreaves}, R.~J., {Gordon}, I.~E., {Rey}, M., {et~al.} 2020, \apjs, 247, 55, \dodoi{10.3847/1538-4365/ab7a1a}

\bibitem[{{Hinkel} {et~al.}(2014){Hinkel}, {Timmes}, {Young}, {Pagano}, \& {Turnbull}}]{hypatia}
{Hinkel}, N.~R., {Timmes}, F.~X., {Young}, P.~A., {Pagano}, M.~D., \& {Turnbull}, M.~C. 2014, \aj, 148, 54, \dodoi{10.1088/0004-6256/148/3/54}

\bibitem[{{Horstman} {et~al.}(2024){Horstman}, {Ruffio}, {Batygin}, {Mawet}, {Baker}, {Hsu}, {Wang}, {Wang}, {Blunt}, {Xuan}, {Xin}, {Liberman}, {Agrawal}, {Konopacky}, {Blake}, {Do {\'O}}, {Bartos}, {Bond}, {Calvin}, {Cetre}, {Delorme}, {Doppmann}, {Echeverri}, {Finnerty}, {Fitzgerald}, {Jovanovic}, {L{\'o}pez}, {Martin}, {Morris}, {Pezzato}, {Ruane}, {Sappey}, {Schofield}, {Skemer}, {Venenciano}, {Wallace}, {Wallack}, \& {Wizinowich}}]{horstman2024}
{Horstman}, K., {Ruffio}, J.-B., {Batygin}, K., {et~al.} 2024, \aj, 168, 175, \dodoi{10.3847/1538-3881/ad73d8}

\bibitem[{{Husser} {et~al.}(2013){Husser}, {Wende-von Berg}, {Dreizler}, {Homeier}, {Reiners}, {Barman}, \& {Hauschildt}}]{phoenix}
{Husser}, T.~O., {Wende-von Berg}, S., {Dreizler}, S., {et~al.} 2013, \aap, 553, A6, \dodoi{10.1051/0004-6361/201219058}

\bibitem[{{Iro} \& {Deming}(2010)}]{iro2010}
{Iro}, N., \& {Deming}, L.~D. 2010, \apj, 712, 218, \dodoi{10.1088/0004-637X/712/1/218}

\bibitem[{{Jovanovic} {et~al.}(2025){Jovanovic}, {Echeverri}, {Delorme}, {Finnerty}, {Schofield}, {Wang}, {Xin}, {Xuan}, {Wallacee}, {Mawet}, {Sanghi}, {Baker}, {Bartos}, {Bond}, {Calvin}, {Cetre}, {Doppmann}, {Fitzgerald}, {Fucik}, {Gao}, {Ge}, {Guthery}, {Horstman}, {Hsud}, {Liberman}, {Leifer}, {Lilley}, {Lopez}, {Marin}, {Martin}, {Mennesson}, {Morris}, {Nash}, {Pezzato}, {Porter}, {Roberts}, {Ruane}, {Ruffio}, {Sappey}, {Serabyn}, {Shen}, {Skemer}, {Wang}, {Wetherell}, {Wizinowich}, {Salama}, {Chambouleyron}, {Jensen-Clem}, \& {Beichman}}]{jovanovic2025}
{Jovanovic}, N., {Echeverri}, D., {Delorme}, J.-R., {et~al.} 2025, arXiv e-prints, arXiv:2502.01863, \dodoi{10.48550/arXiv.2502.01863}

\bibitem[{{Kesseli} {et~al.}(2022){Kesseli}, {Snellen}, {Casasayas-Barris}, {Molli{\`e}re}, \& {S{\'a}nchez-L{\'o}pez}}]{kesseli2022}
{Kesseli}, A.~Y., {Snellen}, I.~A.~G., {Casasayas-Barris}, N., {Molli{\`e}re}, P., \& {S{\'a}nchez-L{\'o}pez}, A. 2022, \aj, 163, 107, \dodoi{10.3847/1538-3881/ac4336}

\bibitem[{{Laughlin} {et~al.}(2009){Laughlin}, {Deming}, {Langton}, {Kasen}, {Vogt}, {Butler}, {Rivera}, \& {Meschiari}}]{laughlin2009}
{Laughlin}, G., {Deming}, D., {Langton}, J., {et~al.} 2009, \nat, 457, 562, \dodoi{10.1038/nature07649}

\bibitem[{{Lei} \& {Molli{\`e}re}(2024)}]{lei2024}
{Lei}, E., \& {Molli{\`e}re}, P. 2024, arXiv e-prints, arXiv:2410.21364, \dodoi{10.48550/arXiv.2410.21364}

\bibitem[{{Lesjak} {et~al.}(2023){Lesjak}, {Nortmann}, {Yan}, {Cont}, {Reiners}, {Piskunov}, {Hatzes}, {Boldt-Christmas}, {Czesla}, {Heiter}, {Kochukhov}, {Lavail}, {Nagel}, {Rains}, {Rengel}, {Rodler}, {Seemann}, \& {Shulyak}}]{lesjak2023}
{Lesjak}, F., {Nortmann}, L., {Yan}, F., {et~al.} 2023, \aap, 678, A23, \dodoi{10.1051/0004-6361/202347151}

\bibitem[{{Lewis} {et~al.}(2017){Lewis}, {Parmentier}, {Kataria}, {de Wit}, {Showman}, {Fortney}, \& {Marley}}]{lewis2017}
{Lewis}, N.~K., {Parmentier}, V., {Kataria}, T., {et~al.} 2017, arXiv e-prints, arXiv:1706.00466, \dodoi{10.48550/arXiv.1706.00466}

\bibitem[{{Line} {et~al.}(2021){Line}, {Brogi}, {Bean}, {Gandhi}, {Zalesky}, {Parmentier}, {Smith}, {Mace}, {Mansfield}, {Kempton}, {Fortney}, {Shkolnik}, {Patience}, {Rauscher}, {D{\'e}sert}, \& {Wardenier}}]{line2021}
{Line}, M.~R., {Brogi}, M., {Bean}, J.~L., {et~al.} 2021, \nat, 598, 580, \dodoi{10.1038/s41586-021-03912-6}

\bibitem[{{L{\'o}pez} {et~al.}(2020){L{\'o}pez}, {Hoffman}, {Doppmann}, {Fitzgerald}, {Johnson}, {Kassis}, {Lanclos}, {Lyke}, {Martin}, {McLean}, {Sohn}, \& {Weiss}}]{nirspecupgrade2}
{L{\'o}pez}, R.~A., {Hoffman}, E.~B., {Doppmann}, G., {et~al.} 2020, in Society of Photo-Optical Instrumentation Engineers (SPIE) Conference Series, Vol. 11447, Society of Photo-Optical Instrumentation Engineers (SPIE) Conference Series, 114476B, \dodoi{10.1117/12.2563075}

\bibitem[{{Martin} {et~al.}(2018){Martin}, {Fitzgerald}, {McLean}, {Doppmann}, {Kassis}, {Aliado}, {Canfield}, {Johnson}, {Kress}, {Lanclos}, {Magnone}, {Sohn}, {Wang}, \& {Weiss}}]{nirspecupgrade}
{Martin}, E.~C., {Fitzgerald}, M.~P., {McLean}, I.~S., {et~al.} 2018, in Society of Photo-Optical Instrumentation Engineers (SPIE) Conference Series, Vol. 10702, Ground-based and Airborne Instrumentation for Astronomy VII, ed. C.~J. {Evans}, L.~{Simard}, \& H.~{Takami}, 107020A, \dodoi{10.1117/12.2312266}

\bibitem[{Mawet {et~al.}(2019)Mawet, Fitzgerald, Konopacky, Beichman, Jovanovic, Dekany, Hover, Chisholm, Ciardi, Artigau, Banyal, Beatty, Benneke, Blake, Burgasser, Canalizo, Chen, Do, Doppmann, Doyon, Dressing, Fang, Greene, Hillenbrand, Howard, Kane, Kataria, Kempton, Knutson, Kotani, Lafreniere, Liu, Nishiyama, Pandey, Plavchan, Prato, Rajaguru, Robertson, Salyk, Sato, Schlawin, Sengupta, Sivarani, Skidmore, Tamura, Terada, Vasisht, Wang, \& Zhang}]{hispec}
Mawet, D., Fitzgerald, M., Konopacky, Q., {et~al.} 2019, arXiv e-prints, \dodoi{10.48550/ARXIV.1908.03623}

\bibitem[{{Mayorga} {et~al.}(2021){Mayorga}, {Robinson}, {Marley}, {May}, \& {Stevenson}}]{mayorga2021}
{Mayorga}, L.~C., {Robinson}, T.~D., {Marley}, M.~S., {May}, E.~M., \& {Stevenson}, K.~B. 2021, \apj, 915, 41, \dodoi{10.3847/1538-4357/abff50}

\bibitem[{{McLean} {et~al.}(1998){McLean}, {Becklin}, {Bendiksen}, {Brims}, {Canfield}, {Figer}, {Graham}, {Hare}, {Lacayanga}, {Larkin}, {Larson}, {Levenson}, {Magnone}, {Teplitz}, \& {Wong}}]{nirspec}
{McLean}, I.~S., {Becklin}, E.~E., {Bendiksen}, O., {et~al.} 1998, in Society of Photo-Optical Instrumentation Engineers (SPIE) Conference Series, Vol. 3354, Infrared Astronomical Instrumentation, ed. A.~M. {Fowler}, 566--578, \dodoi{10.1117/12.317283}

\bibitem[{{Molli{\`e}re} {et~al.}(2019){Molli{\`e}re}, {Wardenier}, {van Boekel}, {Henning}, {Molaverdikhani}, \& {Snellen}}]{prt:2019}
{Molli{\`e}re}, P., {Wardenier}, J.~P., {van Boekel}, R., {et~al.} 2019, \aap, 627, A67, \dodoi{10.1051/0004-6361/201935470}

\bibitem[{{Molli{\`e}re} {et~al.}(2020){Molli{\`e}re}, {Stolker}, {Lacour}, {Otten}, {Shangguan}, {Charnay}, {Molyarova}, {Nowak}, {Henning}, {Marleau}, {Semenov}, {van Dishoeck}, {Eisenhauer}, {Garcia}, {Garcia Lopez}, {Girard}, {Greenbaum}, {Hinkley}, {Kervella}, {Kreidberg}, {Maire}, {Nasedkin}, {Pueyo}, {Snellen}, {Vigan}, {Wang}, {de Zeeuw}, \& {Zurlo}}]{prt:2020}
{Molli{\`e}re}, P., {Stolker}, T., {Lacour}, S., {et~al.} 2020, \aap, 640, A131, \dodoi{10.1051/0004-6361/202038325}

\bibitem[{{Naef} {et~al.}(2001){Naef}, {Latham}, {Mayor}, {Mazeh}, {Beuzit}, {Drukier}, {Perrier-Bellet}, {Queloz}, {Sivan}, {Torres}, {Udry}, \& {Zucker}}]{naef2001}
{Naef}, D., {Latham}, D.~W., {Mayor}, M., {et~al.} 2001, \aap, 375, L27, \dodoi{10.1051/0004-6361:20010853}

\bibitem[{{NASA Exoplanet Science Institute}(2020)}]{10.26133/nea12}
{NASA Exoplanet Science Institute}. 2020, Planetary Systems Table,  IPAC, \dodoi{10.26133/NEA12}

\bibitem[{{NASA Exoplanet Science Institute}(2022)}]{10.26134/exofop5}
---. 2022, Exoplanet Follow-up Observing Program Web Service,  IPAC, \dodoi{10.26134/EXOFOP5}

\bibitem[{Nasedkin {et~al.}(2024)Nasedkin, Mollière, \& Blain}]{Nasedkin2024}
Nasedkin, E., Mollière, P., \& Blain, D. 2024, Journal of Open Source Software, 9, 5875, \dodoi{10.21105/joss.05875}

\bibitem[{{Pai Asnodkar} {et~al.}(2022){Pai Asnodkar}, {Wang}, {Eastman}, {Cauley}, {Gaudi}, {Ilyin}, \& {Strassmeier}}]{pai2022}
{Pai Asnodkar}, A., {Wang}, J., {Eastman}, J.~D., {et~al.} 2022, \aj, 163, 155, \dodoi{10.3847/1538-3881/ac51d2}

\bibitem[{{Pearson} {et~al.}(2022){Pearson}, {Beichman}, {Fulton}, {Esposito}, {Zellem}, {Ciardi}, {Rolfness}, {Engelke}, {Fatahi}, {Zimmerman-Brachman}, {Avsar}, {Bhalerao}, {Boyce}, {Bretton}, {Burnett}, {Burt}, {Cynamon}, {Fowler}, {Gallego}, {Gomez}, {Guillet}, {Hilburn}, {Jongen}, {Kataria}, {Kokori}, {Kumar}, {Kuossari}, {Lekkas}, {Marchini}, {Meneghelli}, {Ngeow}, {Primm}, {Samantaray}, {Shimizu}, {Silvis}, {Sienkiewicz}, {Swain}, {Tan}, {Tock}, {Wagner}, \& {W{\"u}nsche}}]{pearson2022}
{Pearson}, K.~A., {Beichman}, C., {Fulton}, B.~J., {et~al.} 2022, \aj, 164, 178, \dodoi{10.3847/1538-3881/ac8dee}

\bibitem[{{Polyansky} {et~al.}(2018){Polyansky}, {Kyuberis}, {Zobov}, {Tennyson}, {Yurchenko}, \& {Lodi}}]{polyansky2018}
{Polyansky}, O.~L., {Kyuberis}, A.~A., {Zobov}, N.~F., {et~al.} 2018, \mnras, 480, 2597, \dodoi{10.1093/mnras/sty1877}

\bibitem[{{Rosenthal} {et~al.}(2021){Rosenthal}, {Fulton}, {Hirsch}, {Isaacson}, {Howard}, {Dedrick}, {Sherstyuk}, {Blunt}, {Petigura}, {Knutson}, {Behmard}, {Chontos}, {Crepp}, {Crossfield}, {Dalba}, {Fischer}, {Henry}, {Kane}, {Kosiarek}, {Marcy}, {Rubenzahl}, {Weiss}, \& {Wright}}]{rosenthal2021}
{Rosenthal}, L.~J., {Fulton}, B.~J., {Hirsch}, L.~A., {et~al.} 2021, \apjs, 255, 8, \dodoi{10.3847/1538-4365/abe23c}

\bibitem[{{Rothman} {et~al.}(2013){Rothman}, {Gordon}, {Babikov}, {Barbe}, {Chris Benner}, {Bernath}, {Birk}, {Bizzocchi}, {Boudon}, {Brown}, {Campargue}, {Chance}, {Cohen}, {Coudert}, {Devi}, {Drouin}, {Fayt}, {Flaud}, {Gamache}, {Harrison}, {Hartmann}, {Hill}, {Hodges}, {Jacquemart}, {Jolly}, {Lamouroux}, {Le Roy}, {Li}, {Long}, {Lyulin}, {Mackie}, {Massie}, {Mikhailenko}, {M{\"u}ller}, {Naumenko}, {Nikitin}, {Orphal}, {Perevalov}, {Perrin}, {Polovtseva}, {Richard}, {Smith}, {Starikova}, {Sung}, {Tashkun}, {Tennyson}, {Toon}, {Tyuterev}, \& {Wagner}}]{rothman2013}
{Rothman}, L.~S., {Gordon}, I.~E., {Babikov}, Y., {et~al.} 2013, \jqsrt, 130, 4, \dodoi{10.1016/j.jqsrt.2013.07.002}

\bibitem[{{Sikora} {et~al.}(2025){Sikora}, {Rowe}, {Splinter}, {Barat}, {Dang}, {Cowan}, {Barclay}, {Col{\'o}n}, {D{\'e}sert}, {Kane}, {Llama}, {Shivkumar}, {Stassun}, \& {Quintana}}]{sikora2024}
{Sikora}, J.~T., {Rowe}, J.~F., {Splinter}, J., {et~al.} 2025, \aj, 170, 105, \dodoi{10.3847/1538-3881/addfda}

\bibitem[{Skilling(2004)}]{skilling2004}
Skilling, J. 2004, AIP Conference Proceedings, 735, 395, \dodoi{10.1063/1.1835238}

\bibitem[{{Smith} {et~al.}(2024){Smith}, {Sanchez}, {Line}, {Rauscher}, {Mansfield}, {Kempton}, {Savel}, {Wardenier}, {Pino}, {Bean}, {Beltz}, {Panwar}, {Brogi}, {Malsky}, {Fortney}, {D{\'e}sert}, {Pelletier}, {Parmentier}, {Kanumalla}, {Welbanks}, {Meyer}, \& {Monnier}}]{smith2024}
{Smith}, P. C.~B., {Sanchez}, J.~A., {Line}, M.~R., {et~al.} 2024, \aj, 168, 293, \dodoi{10.3847/1538-3881/ad8574}

\bibitem[{{Snellen}(2025)}]{snellen2025}
{Snellen}, I. 2025, arXiv e-prints, arXiv:2505.08926, \dodoi{10.48550/arXiv.2505.08926}

\bibitem[{{Snellen} {et~al.}(2010){Snellen}, {de Kok}, {de Mooij}, \& {Albrecht}}]{snellen2010}
{Snellen}, I. A.~G., {de Kok}, R.~J., {de Mooij}, E. J.~W., \& {Albrecht}, S. 2010, \nat, 465, 1049, \dodoi{10.1038/nature09111}

\bibitem[{{Tsai} {et~al.}(2023){Tsai}, {Lee}, {Powell}, {Gao}, {Zhang}, {Moses}, {H{\'e}brard}, {Venot}, {Parmentier}, {Jordan}, {Hu}, {Alam}, {Alderson}, {Batalha}, {Bean}, {Benneke}, {Bierson}, {Brady}, {Carone}, {Carter}, {Chubb}, {Inglis}, {Leconte}, {Line}, {L{\'o}pez-Morales}, {Miguel}, {Molaverdikhani}, {Rustamkulov}, {Sing}, {Stevenson}, {Wakeford}, {Yang}, {Aggarwal}, {Baeyens}, {Barat}, {de Val-Borro}, {Daylan}, {Fortney}, {France}, {Goyal}, {Grant}, {Kirk}, {Kreidberg}, {Louca}, {Moran}, {Mukherjee}, {Nasedkin}, {Ohno}, {Rackham}, {Redfield}, {Taylor}, {Tremblin}, {Visscher}, {Wallack}, {Welbanks}, {Youngblood}, {Ahrer}, {Batalha}, {Behr}, {Berta-Thompson}, {Blecic}, {Casewell}, {Crossfield}, {Crouzet}, {Cubillos}, {Decin}, {D{\'e}sert}, {Feinstein}, {Gibson}, {Harrington}, {Heng}, {Henning}, {Kempton}, {Krick}, {Lagage}, {Lendl}, {Lothringer}, {Mansfield}, {Mayne}, {Mikal-Evans}, {Palle}, {Schlawin}, {Shorttle}, {Wheatley}, \& {Yurchenko}}]{tsai2023}
{Tsai}, S.-M., {Lee}, E. K.~H., {Powell}, D., {et~al.} 2023, \nat, 617, 483, \dodoi{10.1038/s41586-023-05902-2}

\bibitem[{{Villanueva} {et~al.}(2018){Villanueva}, {Smith}, {Protopapa}, {Faggi}, \& {Mandell}}]{psg}
{Villanueva}, G.~L., {Smith}, M.~D., {Protopapa}, S., {Faggi}, S., \& {Mandell}, A.~M. 2018, \jqsrt, 217, 86, \dodoi{10.1016/j.jqsrt.2018.05.023}

\bibitem[{{Wardenier} {et~al.}(2025){Wardenier}, {Parmentier}, {Lee}, \& {Line}}]{wardenier2025}
{Wardenier}, J.~P., {Parmentier}, V., {Lee}, E. K.~H., \& {Line}, M.~R. 2025, \apj, 986, 63, \dodoi{10.3847/1538-4357/add341}

\bibitem[{{Wardenier} {et~al.}(2021){Wardenier}, {Parmentier}, {Lee}, {Line}, \& {Gharib-Nezhad}}]{wardenier2021}
{Wardenier}, J.~P., {Parmentier}, V., {Lee}, E. K.~H., {Line}, M.~R., \& {Gharib-Nezhad}, E. 2021, \mnras, 506, 1258, \dodoi{10.1093/mnras/stab1797}

\bibitem[{{Wardenier} {et~al.}(2023){Wardenier}, {Parmentier}, {Line}, \& {Lee}}]{wardenier2023}
{Wardenier}, J.~P., {Parmentier}, V., {Line}, M.~R., \& {Lee}, E. K.~H. 2023, \mnras, 525, 4942, \dodoi{10.1093/mnras/stad2586}

\bibitem[{{Welbanks} {et~al.}(2025){Welbanks}, {Nixon}, {McGill}, {Tilke}, {Wiser}, {Rotman}, {Mukherjee}, {Feinstein}, {Line}, {Seager}, {Beatty}, {Seligman}, {Parmentier}, \& {Sing}}]{welbanks2025}
{Welbanks}, L., {Nixon}, M.~C., {McGill}, P., {et~al.} 2025, arXiv e-prints, arXiv:2504.21788, \dodoi{10.48550/arXiv.2504.21788}

\end{thebibliography}
\bibliographystyle{aasjournal}



\end{document}